\documentstyle[12pt, twoside, epsf, here, psfig]{article}
\newcounter{tony}
\newcommand{\bb}{\mbox{\boldmath{$b$}}}

\newcommand{\bn}{\mbox{\boldmath{$n$}}}

\newcommand{\bu}{\mbox{\boldmath{$u$}}}
\newcommand{\bv}{\mbox{\boldmath{$v$}}}
\newcommand{\bw}{\mbox{\boldmath{$w$}}}
\newcommand{\bx}{\mbox{\boldmath{$x$}}}

\newcommand{\bA}{\mbox{\boldmath{$A$}}}
\newcommand{\bB}{\mbox{\boldmath{$B$}}}

\newcommand{\bD}{\mbox{\boldmath{$D$}}}

\newcommand{\bF}{\mbox{\boldmath{$F$}}}

\newcommand{\bI}{\mbox{\boldmath{$I$}}}

\newcommand{\bN}{\mbox{\boldmath{$N$}}}

\newcommand{\bP}{\mbox{\boldmath{$P$}}}

\newcommand{\blambda}{\mbox{\boldmath{$\lambda$}}}

\newcommand{\bsigma}{\mbox{\boldmath{$\sigma$}}}

\newcommand{\beq}{\begin{equation}}
\newcommand{\eeq}[1]{\label{eq:#1}\end{equation}}

\newcommand{\eqref}[1]{(\ref{eq:#1})}

      \newcommand{\beqn}{\begin{equation}}
      \newcommand{\eeqn}{\end{equation}}
      \newcommand{\beqna}{\begin{eqnarray}}
      \newcommand{\eeqna}{\end{eqnarray}}

\newtheorem{lemma}{\sc Lemma}
\newtheorem{theorem}{\sc Theorem}
\newtheorem{inequality}{\sc Inequality}
\newtheorem{assumption}{\sc Assumption}

\title{The Energetic Implications of the Time Discretisation in Implementations of the A.L.E. Equations}

\author{S. J. Childs \\ \\ {\small\em Department of Pure and Applied
Mathematics, Rhodes University, Grahamstown,} \\ {\small\em 6140, South
Africa}}

\renewcommand{\thefootnote}{\fnsymbol{footnote}}
\date{}       

\begin{document}

\maketitle
\renewcommand{\thefootnote}{\arabic{footnote}}

\begin{abstract}
\noindent {\em A class of A.L.E. time discretisations which inherit
key energetic properties (nonlinear dissipation in the absence of forcing and long--term stability under conditions of time dependent loading), irrespective of the time increment employed, is established in this work. These properties are intrinsic to real flows and the conventional Navier--Stokes equations. \\

\noindent A description of an incompressible, Newtonian fluid, which
reconciles the differences between the various schools of A.L.E.
thought in the literature is derived for the purposes of this
investigation. The issue of whether these equations automatically
inherit the afore mentioned energetic properties must first be
resolved. In this way natural notions of nonlinear, exponential--type
dissipation in the absence of forcing and long--term stability under
conditions of time dependent loading are also formulated. \\

\noindent The findings of this analysis have profound
consequences for the use of certain classes of finite difference
schemes in the context of deforming references. It is significant
that many algorithms presently in use do not automatically inherit
the fundamental qualitative features of the dynamics. \\

\noindent The main conclusions are drawn on in the simulation of a
driven cavity flow, a driven cavity flow with various, included rigid
bodies, a die--swell problem, and a Stokes second order wave. The
improved, second order accuracy of a new scheme for the linearised
approximation of the convective term is proved for the purposes of
these simulations. A somewhat novel method to generate finite element
meshes automatically about included rigid bodies, and which involves
finite element mappings, is also described.} 
\end{abstract}


\section{Introduction}

This work focusses on establishing a class of A.L.E. time
discretisations which inherit key energetic properties (nonlinear,
exponential--type dissipation in the absence of forcing and
long--term stability under conditions of time dependent loading)
irrespective of the time increment employed. The findings of this
analysis have profound consequences for the use of certain classes of
difference schemes in the context of deforming references. It is
significant that many algorithms presently in use do not
automatically inherit the fundamental qualitative features of the
dynamics.

Descriptions of fluid motion are conventionally based on the
principles of conservation of mass and linear momentum. One might
hope that all such descriptions would accordingly exhibit the afore
mentioned, key energetic properties consistant with the principle of
energy conservation. These properties are intrinsic to real flows and
the conventional, Eulerian Navier--Stokes equations.

A description of an incompressible, Newtonian fluid, which reconciles
the differences between the so--called arbitrary Lagrangian Eulerian
(A.L.E.) formulation of {\sc Hughes, Liu} and {\sc Zimmerman}
\cite{h:1} (deformation gradients absent) and that of {\sc
Soulaimani, Fortin, Dhatt} and {\sc Ouellet} \cite{f:1} (deformation
gradients present, but use is problematic), is derived for the
purposes of this investigation. The implications of the resulting
description are investigated in the context of energy conservation in
a similar, but broader, approach to that taken by others (eg. {\sc
Simo} and {\sc Armero} \cite{s:1}) for the conventional, Eulerian
Navier--Stokes equations. 

The main conclusions of this work rely on a new inequality and a
number of lemmas, the proofs of which are listed in an appendix at
the end of the paper. The new inequality is used in place of where
the Poincar\'{e}--Friedrichs inequality might otherwise have limited
the analysis. The lemmas are mainly concerned with the new convective
term. This analysis is extended in that non--zero boundaries,
so--called free boundaries and time--dependent loads are considered.

The resulting theory is used in the simulation of a driven cavity
flow, a driven cavity flow with various, included rigid bodies, a
die--swell problem, and a Stokes second order wave. A new scheme for
the linearised approximation of the convective term is proposed and
the improved, second order accuracy of this scheme is proved for the
purposes of these simulations. A somewhat novel method to generate
finite element meshes automatically about included rigid bodies, and
which involves finite element mappings, is also described.

\section{A Completely General Reference} \label{130}

The implementation of most numerical time integration schemes would
be problematic were a conventional
Eulerian\footnotemark[1]\footnotetext[1]{{\sc Eulerian} or {\sc
spatial} descriptions are in terms of fields defined over the current
configuration.} description of fluid motion to be used in instances
involving deforming domains. The reason is that most numerical time
integration schemes require successive function evaluation at fixed
spatial locations (the exception being the finite element with
respect to time approach of {\sc Tezduyar, Behr} and {\sc Liou}
\cite{t:1}). On the other hand meshes rapidly snarl when purely
Lagrangian\footnotemark[2]\footnotetext[2]{{\sc Lagrangian} or {\sc
material} descriptions are made in terms of fields defined over a
reference (a material reference) configuration.} descriptions are
used. It is for these reasons that a completely general reference
description is usually resorted to.

Eulerian and Lagrangian references are just two, specific examples of
an unlimited number of configurations over which to define fields
used to describe the dynamics of deforming continua. They are both
special cases of a more general reference description, a description
in which the referential configuration is deformed at will and which
is the focus of this investigation. A deforming finite element mesh
would be a good example of just such a deforming reference in
practice. The transformation to the completely general reference
involves coordinates where used as spatial variables only and the
resultant description is therefore inertial in the same way as
Lagrangian descriptions are.

\subsection{Domains, Mappings and a Notation}

Consider a material body which occupies a domain $\Omega$ at time $t$.
The material domain, $\Omega_0$, is that corresponding to time $t =
t_0$ (the reference time, $t_0$, is conventionally, but not always,
zero). A third configuration, ${\tilde \Omega}$, which is chosen
arbitrarily is also defined for the purposes of this work. The three
domains are related in the sense that points in one domain may be
obtained as one--to--one invertible maps from points in another.
\begin{figure}[h]
\begin{center}
\mbox{\epsfbox{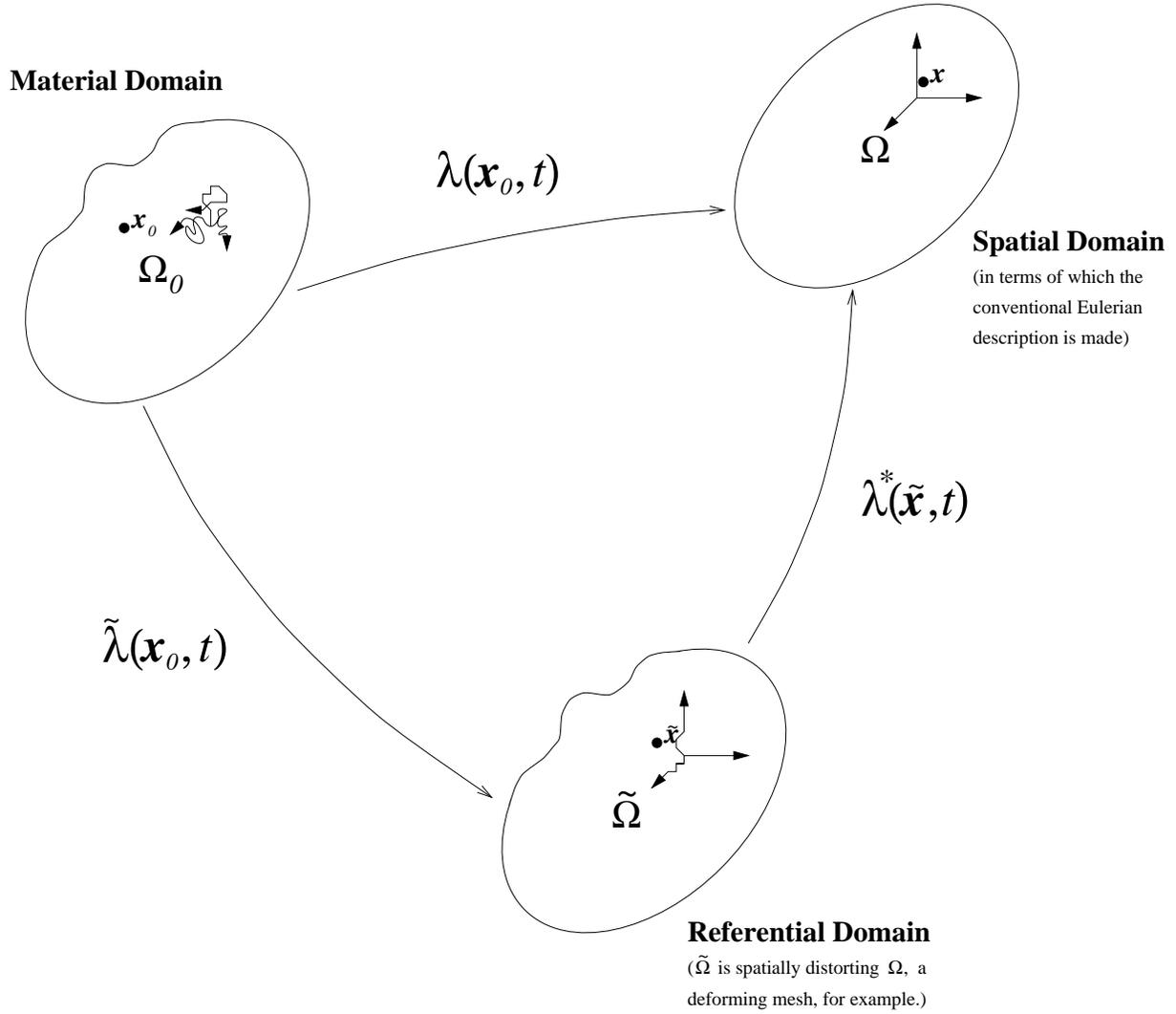}}
\end{center}
\caption{Schematic Diagram of Domains and Mappings Used in a
completely general reference Description} \label{91}
\end{figure}

For any general function $f({\bx},t)$, a function, $\tilde
f(\tilde{\bx},t) \ \equiv \ f({\blambda}^{\ast}(\tilde{\bx},t),t)$,
can be defined in terms of the domains and one--to--one, invertible
mappings illustrated in Fig. \ref{91}. Similarly, $f_0({\bx}_0 ,t)
\ \equiv \ f({\blambda}({\bx}_0 ,t),t)$ can be defined. This notation
can be generalised for the component--wise definition of higher order
tensors. The key to understanding much of this work lies possibly in
adopting a component-wise defined notation.

In contrast to the function notation just established, the definition
of the operators ${\tilde \nabla}$ and $\widetilde {\mbox{div}}$ is
not based on $\nabla$ and $\mbox{div}$. They are instead the
referential counterparts, that is
\[ 
{\tilde \nabla} = \frac{\partial}{\partial {\tilde {\bx}}}
\hspace{10mm} \mbox{and} \hspace{10mm} {\widetilde {\mbox{div}}} =
\frac{\partial}{\partial {\tilde x}_1} + \frac{\partial}{\partial
{\tilde x}_2} + \frac{\partial}{\partial {\tilde x}_3}.
\] 
The notation ${\bA}:{\bB}$ is used to denote the matrix inner product
$A_{ij}B_{ij}$ throughout this work, $\left< \ \cdot \ , \ \cdot
\ \right>_{L^2( \ \cdot \ )}$ denotes the $L^2$ inner product and
$\left|\left| \ \cdot \ \right|\right|_{L^2( \ \cdot \ )}$ the $L^2$
norm.

\subsection{Some General Results for Functions Defined on the Three
Domains}

Three important results are necessary for the derivation of the
completely general reference description and these are presented
below.

\subsubsection*{The Material Derivative in Terms of a Completely
General Reference} \label{2}

The material derivative of any vector field ${\tilde {\bv}}$ in terms
of a completely general, reference is
\begin{eqnarray} \label{25} 
\frac{\partial \tilde{\bv}}{\partial t} + {{\tilde \nabla} \tilde
{\bv}} \left[ {\tilde {\bF}}^{-1} (\tilde{\bv} - \tilde{\bv}^{ref})
\right].
\end{eqnarray}      
where $\tilde{\bv}^{ref}$ is the velocity of the reference deformation,
and $\tilde {\bF}$ is the deformation gradient given by
\[
{\tilde {\bF}}({\tilde {\bx}}) = \frac{\partial {\blambda}^*}{\partial
{\tilde {\bx}}}.
\]
This result is demonstrated in Appendix II.

\subsubsection*{An Element of Area in Terms of a Distorting Reference} 
The second important result can be recalled from general continuum
mechanics. Consider an element of area, size $dA$, with an outward unit
normal ${\bn}$. Then
\begin{eqnarray} \label{27}
{\bn} dA = {\tilde{\bF}}^{-t}\tilde{\bN} {\tilde J}
d\tilde{A} 
\end{eqnarray} 
where $d{\tilde A}$ and $\tilde{\bN}$ denote the respective analogous
size and outward unit normal of this element of area in the
referential configuration and ${\tilde J} = \det{\tilde {\bF}}$. This
result is demonstrated in most popular textbooks on continuum
mechanics (eg. {\sc Lai}, {\sc Rubin} and {\sc Krempl} \cite{lrk:1}).

{\sc Remark:} Notice that ${\tilde {\bN}}$ is the single exception to
the \ ${\tilde {}}$ \ quantities devised in this work. ${\tilde
{\bN}}$ is the surface normal perceived in a completely general
reference and the components of ${\bn}$ and ${\tilde {\bN}}$ need
have nothing in common (${\tilde {\bN}}$ is not ${\tilde {\bn}}$).

\subsubsection*{The Kinematic Relation ${\dot {\cal J}}_0 = {\cal J}_0
\mathop{\rm div}{\bv}$} \label{5}

The material derivative of the Jacobian ${\cal J}_0$ is given by the
relation
\[
{\dot {\cal J}}_0 = {\cal J}_0 \mathop{\rm div}{\bv}
\]
where ${\cal J}_0$ is defined as follows,
\[
{\cal J}_0 \equiv \det \left\{ \frac{\partial {\blambda}}{\partial
{\bx}_0} \right\}.
\]
This result is demonstrated in most popular textbooks on continuum
mechanics (eg. {\sc Marsden} and {\sc Hughes} \cite{mh:1}).

\subsection{Derivation of the Completely General Equation}

One way in which to derive a completely general reference
description of an incompressible, Newtonian fluid is to start with
the balance laws in global (integral) form, and to make the necessary
substitutions in these integrals. The desired numerical
implementation (similar to the conventional Navier--Stokes one which
has been thoroughly investigated and found to be stable) is then
obtained.

\subsubsection*{Conservation of Mass}

Let $\Omega(t)$ be an arbitrary sub--volume of material. The principle
of conservation of mass states that
\begin{eqnarray} \label{92}
\frac{d}{dt} \int_{\Omega(t)} \rho {d \Omega} &=& 0 \hspace{10mm}
\mbox{\it (rate of change of mass with time }= \ \mbox{\it 0)}
\nonumber \\ 
& & \nonumber \\
\frac{d}{dt} \int_{\Omega_0} {{\rho}_0} {\cal J}_0 {d \Omega_0} &=& 0
\hspace{10mm} \mbox{\it (reformulating in terms of the material}
\nonumber \\
& & \hspace{13mm} \mbox{\it configuration, } \Omega_0 \mbox{\it .)}
\nonumber \\
\int_{\Omega_0} \frac{\partial}{\partial t}  \left\{ {{\rho}_0} {\cal
J}_0 \right\} {d \Omega_0} &=& 0 \hspace{10mm} \mbox{\it (since limits
are not time dependent in} \nonumber \\
& & \hspace{13mm} \mbox{\it the material configuration.)} \\
\int_{\Omega_0} \left( \rho_0 {\dot {\cal J}}_0 \ + \ {\dot \rho}_0
{\cal J}_0 \right) d \Omega_0 &=& 0 \hspace{10mm} \mbox{\it (by the
chain rule)} \nonumber \\
& & \nonumber \\
\int_{\Omega(t)} \left( \dot{\rho} \ + \ {\rho \mathop{\rm div}{\bv} }
\right) {d \Omega} &=& 0 \hspace{10mm} \mbox{\it (using the kinematic
result } \dot{{\cal J}_0} = {\cal J}_0 \mbox{div}\, {\bv} \mbox{\it )} \nonumber
\\
& & \nonumber \\
\int_{{\tilde \Omega}(t)} \left( {\dot {\rho}} \ + \ {\rho}
\frac{\partial \tilde{v}_i}{\partial \tilde{x}_j}\frac{\partial
\tilde{x}_j}{\partial x_i} \right) {\tilde J} {d
{\tilde \Omega}} &=& 0 \hspace{10mm} \mbox{\it (reformulating in terms
of the distorting} \nonumber \\
& & \hspace{13mm} \mbox{\it referential configuration, }{\tilde \Omega}
(t) \mbox{\it .)} \nonumber \\
\Rightarrow \left( {\dot {\rho}} \ + \ {\rho} {\tilde \nabla} {\tilde {\bv}} : {\tilde {\bF}}^{-t} \right) {\tilde J} &=& 0 \hspace{10mm}
\mbox{\it (integrand must be zero since the volume} \nonumber \\
& & \hspace{13mm} \mbox{\it was arbitrary.)} \nonumber 
\end{eqnarray} 
Thus, for a material of constant, non--zero density,
\[
{\tilde \nabla} {\tilde {\bv}} : {\tilde {\bF}}^{-t} = 0 \hspace{7mm}
\mbox{since} \hspace{7mm} {\tilde J} \neq 0
\hspace{12mm} \mbox{\it (mappings are one-to-one and invertible).}
\]
Notice also that equation (\ref{92}) implies 
\begin{eqnarray} \label{93}
\frac{\partial}{\partial t}  \left\{ {{\rho}_0} {\cal J}_0 \right\} &=&
0
\end{eqnarray}
since the volume was arbitrary and the integrand must therefore be
zero.

\subsubsection*{Conservation of Linear Momentum (and Mass)}

The principle of conservation of linear momentum for an arbitrary
volume of material $\Omega(t)$ with boundary $\Gamma(t)$ states that
\begin{eqnarray} \label{28} 
\frac{d}{dt} \int_{\Omega(t)} \rho {\bv} {d \Omega} = \int_{\Omega(t)}
\rho {\bb} {d \Omega} \ + \  \int_{\Gamma(t)} {\bsigma} {\bn} dA
\end{eqnarray}
where ${\rho}$ is density, ${\bb}$  is the body force per unit mass,
${\bsigma}$ is the stress, {\bn}  the outward unit normal to the
boundary and {\bv}  is the velocity. The term on the lefthand side can
be rewritten as follows:
\begin{eqnarray*}
\frac{d}{dt} \int_{\Omega(t)} \rho {\bv}  {d \Omega} &=& \frac{d}{dt}
\int_{\Omega_0} {{\rho}_0} {{\bv} _0} {\cal J}_0 {d \Omega_0} \hspace{10mm}
\mbox{\it (Reformulating in terms of the material} \\
& & \hspace{42mm} \mbox{\it configuration, } \Omega_0\mbox{\it .)} \\ 
&=& \int_{\Omega_0} \frac{\partial}{\partial t}  \left\{ {{\rho}_0}
{{\bv} _0} {\cal J}_0  \right\} {d \Omega_0} \hspace{5mm} \mbox{\it (Since
limits are not time dependent in} \\
& & \hspace{41mm} \mbox{\it the material configuration.)} \\
&=&  \int_{\Omega_0} \left( \frac{\partial {\bv}_0}{\partial t}
{{\rho}_0} {\cal J}_0 \ + \ {{\bv}_0}\frac{\partial}{\partial t} \left\{
{{\rho}_0} {\cal J}_0 \right\} \right) {d \Omega_0} \\
& &  \\
& &  \\
&=& \int_{\Omega(t)} \rho \dot{\bv}  {d \Omega} \hspace{22mm} \mbox{\it
(The second term above is zero as a} \\
& & \hspace{41mm} \mbox{\it consequence of equation (\ref{93}).)} \\
&=& \int_{{\tilde \Omega}(t)} {\rho} {\dot{\tilde {\bv} }} {\tilde J}
{d {\tilde \Omega}}  \hspace{20mm} \mbox{\it (Reformulating in terms of
the dist--} \\
& & \hspace{42mm} \mbox{\it orting referential configuration, } \tilde
\Omega \mbox{\it .)} \\
&=& \int_{{\tilde \Omega}(t)} {\rho} \left( \frac{\partial
\tilde{\bv}}{\partial t} + { {{\tilde \nabla} \tilde {\bv} } } \left[
{\tilde {\bF}}^{-1} (\tilde{\bv}  - \tilde{\bv} ^{ref}) \right]
\right) {\tilde J} {d {\tilde \Omega}} \hspace{5mm}
\mbox{\it (Using result} \\
&& \hspace{79mm} \mbox{\it (\ref{25}) on page \pageref{25})}
\end{eqnarray*} 
where $\dot{\bv}$ denotes the material derivative of ${\bv} $. The
surface integral becomes
\begin{eqnarray*}
\begin{array}{rcll}
\displaystyle \int_{\Gamma(t)} {\bsigma} {\bn}  dA &=& \displaystyle
\int_{\tilde {\Gamma}(t)} {\tilde {\bsigma}} \tilde{\bF} ^{-t} {\tilde
{\bN}} {\tilde J} d{\tilde A} & \mbox{\it (Reformulating in terms of a
distorting} \\
& & & \mbox{\it \ reference using result (\ref{27}) on page
\pageref{27}.)} \\
&=& \displaystyle \int_{{\tilde \Omega}(t)} \mathop{\widetilde {\rm
div}}\, \{ {\tilde {\bsigma}} {\tilde{\bF} ^{-t}} {\tilde J} \} {d
{\tilde {\Omega}} } & \mbox{\it (By the divergence theorem).}
\end{array} 
\end{eqnarray*} 
Finally, the term involving body force becomes
\begin{eqnarray*}
\begin{array}{rcll}
\displaystyle \int_{\Omega(t)} \rho {\bb}  {d \Omega} &=& \displaystyle
\int_{{\tilde \Omega}(t)} {\rho} {\tilde {\bb} } {\tilde J} {d {\tilde
\Omega}} \hspace{18mm} & \mbox{\it (Reformulating in terms of a
distorting} \\
& & & \mbox{\it \ reference.).}   
\end{array}
\end{eqnarray*}
Substituting these expressions into (\ref{28}), remembering that the
volume used in the argument was arbitrary and that the entire integrand
must therefore be zero, the conservation principles of linear momentum
and mass may be written in primitive form as
\begin{eqnarray} \label{29}
{\rho} \left( \frac{\partial {\tilde {\bv} }}{\partial t} + {{{\tilde
\nabla} \tilde {\bv} } {\tilde {\bF} }^{-1} } ({\tilde {\bv} } -
{\tilde {\bv} }^{ref}) \right) {\tilde J} &=& {\rho} {\tilde {\bb} }
{\tilde J} + \mathop{\widetilde {\rm div}}{\tilde {\bP}}
\end{eqnarray}
and
\begin{eqnarray} \label{30}
{{\tilde \nabla} {\tilde {\bv}} : {\tilde {\bF} }^{-t} } \ = \ 0 
\end{eqnarray}
where $\tilde{\bP}$ is the Piola--Kirchoff stress tensor of the first
kind, $\tilde{\bP}  \ = \ \tilde{\bsigma} {\tilde {\bF}}^{-t} {\tilde
J}$. In terms of the constitutive relation, $\bsigma = - p {\bI} + 2
\mu {\bD}$, for a Newtonian fluid,
\[
{\tilde {\bP}} = \left( - p{\bI} + \mu \left[ {\tilde \nabla}{\tilde
{\bv}}{\tilde {\bF}}^{-1} + \left({\tilde \nabla}{\tilde
{\bv}}{\tilde {\bF}}^{-1}\right)^t \right] \right){\tilde {\bF}}^{-t}
{\tilde J} \hspace{8mm} \mbox{since} \hspace{8mm} {\tilde {\bD}} =
\frac{1}{2} \left( {\widetilde {\nabla {\bv}}} + \left(
{\widetilde{\nabla {\bv}}} \right)^t \right).
\]
The derivation of a variational formulation is along similar lines as
that for the Navier-Stokes equations (the purely Eulerian
description). For a fluid of constant density, the variational
formulation
\renewcommand{\thefootnote}{\fnsymbol{footnote}}
\begin{eqnarray} \label{33}
{\rho} \int_{ \tilde \Omega } {\tilde {\bw} } \cdot
\frac{\partial {\tilde {\bv} }}{\partial t} {\tilde J}
{d{\tilde \Omega}} \ + \ {\rho} \int_{ \tilde \Omega } {\tilde
{\bw}} \cdot {{\tilde \nabla} {\tilde {\bv} }} \left[{\tilde {\bF}
}^{-1} ({\tilde {\bv}} - {\tilde {\bv} }^{ref}) \right] {{\tilde J}}{d {\tilde \Omega}} \ = \nonumber \hspace{40mm} & & \\
{\rho} \int_{ \tilde \Omega } {\tilde {\bw} } \cdot {\tilde
{\bb}} {\tilde J} {d {\tilde \Omega}} \ + \ \int_{
\tilde \Omega } {\tilde p} {{{\tilde \nabla} \tilde {\bw} } : {\tilde
{\bF} }^{-t}} {\tilde J} {d {\tilde \Omega}} \ - \ 2
{\mu} \int_{ \tilde \Omega } {\tilde {\bD} }(\tilde {\bw} ) :
{\tilde {\bf D}}(\tilde {\bv}) {\tilde J} {d {\tilde
\Omega}} \nonumber & & \\
+  {\rho} \int_{ \tilde \Gamma } {\tilde {\bw}} {\tilde {\bP}} {\tilde
{\bN}} d {\tilde \Gamma} \hspace{60mm} & & 
\end{eqnarray}
\begin{eqnarray} \label{34}
\int_{\tilde \Omega} {\tilde q} { {{\tilde \nabla} \tilde {\bv} } :
{\tilde {\bF} }^{-t} } {d {\tilde \Omega}} &=& 0
\end{eqnarray}
is obtained, where $\tilde q$ and ${\tilde {\bw}}$ are respectively
the arbitrary pressure and velocity of the variational formulation.
\renewcommand{\thefootnote}{\arabic{footnote}}

Notice that the usual procedure of assigning a value of zero to the
arbitrary velocity, ${\tilde {\bw}}$, at the boundary has not been
followed. The boundary integral in the variational momentum equation
has consequently not been eliminated as is normally done. The reasons
are twofold; firstly problems for which the ensuing investigation is
intended are of a free boundary type and so the solution is not known
there; secondly, a specific function (which cannot arbitrarily be
assigned a value of zero at the boundary) will be substituted for
${\tilde {\bw}}$ in the forthcoming analysis.

\subsection{Reconciling the Different Schools of Thought} \label{1001}

The equations (\ref{29}) and (\ref{30}) are the completely general
referential description of an incompressible, Newtonian fluid. They
reduce to the so--called A.L.E. equations of {\sc Hughes, Liu} and
{\sc Zimmerman} \cite{h:1} for an instant in which spatial and
referential configurations coincide. 

Since the approximate set of equations is broken into a sequence of
discrete time steps in the implementation, one is entitled to choose
a new referential configuration during each time step, should one so
desire. This is what is known as an ``updated'' approach; when each
time step is really a fresh implementation. In the case of time
stepping schemes based about a single instant (eg. the generalised
class of Euler difference schemes to be investigated in Section
\ref{6}) a considerably simplified implementation can be achieved by
an appropriate choice of configurations. Making the choice of a
referential configuration which coincides with the spatial
configuration at the instant about which the time stepping scheme is
based allows the deformation gradient to be omitted from the
approximation altogether (the deformation gradient is identity under
such circumstances). For such implementations (those which require
evaluation about a single point only) no error arises from the use of
the equations cited in {\sc Hughes, Liu} and {\sc Zimmerman}
\cite{h:1},
\begin{eqnarray} 
{\rho} \left( \frac{\partial { {\bv} }}{\partial t} + {{{
\nabla}  {\bv} }  } ({ {\bv} } - { {\bv} }^{ref})
\right) &=& {\rho} { {\bb} } + \mathop{ {\rm
div}}{ {\bsigma}} \label{307} \\
\mathop{ {\rm div}}{ {\bv}} &=& 0. \label{308}
\end{eqnarray}
These equations are not valid for any, arbitrary choice of reference
or if the implementation requires the equation to be evaluated at
more than one point within each time step (eg. a Runge--Kutta or
finite--element--in--time scheme). It is important to remember that
in a discrete context the reference configuration is fixed for the
duration of the entire time increment. Although the referential
configuration is hypothetical and can be chosen arbitrarily for each
time step, once chosen it is static for the duration of the entire
time step. Once the coincidence of configurations is ordained at a
given instant, ${\tilde {\bF}}$ is defined by the reference (mesh)
deformation, both before and after, and must be consistant.

The equations of {\sc Hughes et al.} are an arbitrary Lagrangian
Eulerian (A.L.E.) description in the very true sense under the
circumstances of implementations requiring evaluation about more than
one point within each time step (this is not surprising considering
the equations have their origins in the arbitrarily, either
Lagrangian or Eulerian programmes of {\sc Hirt}, {\sc Amsden} and
{\sc Cook} \cite{h:5}). This fact is further bourne out in observing
that key energetic properties, consistant with the principle of
energy conservation, are not automatically inherited by the equations
of {\sc Hughes et. al.} in the context of more general references.

The momentum equations of {\sc Soulaimani, Fortin, Dhatt} and {\sc
Ouellet} \cite{f:1} are flawed as a result of the mistaken belief that
${\tilde {\bsigma}} {\tilde {\bF}}^{-1} {\tilde J}$ is the
Piola--Kirchoff stress tensor of the first kind (pg. 268 of {\sc
Soulaimani et al.}). Yet another problem is illustrated by rewriting
the conventional incompressibility condition using the chain
rule. The new incompressibility condition which arises is most
certainly
\[
\frac{\partial \tilde{v}_i}{\partial \tilde{x}_j}\frac{\partial
\tilde{x}_j}{\partial x_i} = 0 \hspace{10mm} \mbox{and not}
\hspace{10mm} \frac{\partial \tilde{v}_i}{\partial
\tilde{x}_j}\frac{\partial \tilde{x}_i}{\partial x_j} = 0.
\]
Further errors arising (eg. $\hat J$ omitted in the first term on the
right hand side of the momentum equation, equation (10) on pg. 268 of
{\sc Soulaimani et al.}) make the use of these equations problematic.

There would seem to be no reason why one would wish to define the
deformation about a configuration other than that at the instant about
which the implementation is based (assuming the implementation used is
indeed based about a single point eg. a finite difference) thereby
involving deformation gradients. Resolving the resulting difficulties
associated with the deformation gradients by means of a perturbation
seems unnecessarily complicated in the light of the above reasoning.

\section{Natural Notions of Energy Conservation in Terms of the Completely General Equations} \label{69}

The effect of quantities parameterising reference deformation on key
energetic properties -- nonlinear, exponential--type dissipation in the absence of forcing and long--term stability under conditions of time dependent loading -- is investigated in this section. These properties, of form
\[
K({\bv}) \le K({\bv} \mid_{t_0}) \ e^{-2 {\nu} C t} \hspace{10mm}
\mbox{and} \hspace{10mm} \lim_{t \rightarrow \infty} \sup K({\bv}) \le
\frac{M^2}{2 \nu^2 C^2}
\]
respectively (where $K = \frac{1}{2} {\rho} \left|\left| {\bv}
\right|\right|_{L^2(\Omega)}^2$ is the total kinetic energy), are
intrinsic to real flows and the conventional, Eulerian Navier--Stokes
equations (see {\sc Temam} \cite{Temam:1}, \cite{Temam:2}, {\sc
Constantin} and {\sc Foias} \cite{Constantin:1} and {\sc Simo} and
{\sc Armero} \cite{s:1} in this regard). The effect of ${\tilde
{\bv}}^{ref}$ on the afore mentioned aspects of conservation of the
quantity
\[
{\tilde K}({\tilde {\bv}}) \equiv \frac{1}{2} {\rho} \left|\left| {\tilde {\bv}} {\tilde J}^{\frac{1}{2}}  \right|\right|_{L^2({\tilde \Omega})}^2
\]
is essentially what is being investigated, with a view to establishing
a set of conditions under which the discrete approximation can
reasonably be expected to inherit these self--same energetic
properties.

One might anticipate key energetic properties to be manifest only in
instances involving a fixed contributing mass of material, whether its
boundaries be dynamic, or not. An analysis of this nature only makes
sense in the context of a constant volume of incompressible fluid. 

Inequalities of the Poincar\'e-Friedrichs type are a key feature of
any stability analysis of this nature. Gradient containing $L^2$
terms need to be re--expressed in terms of energy. In the case of a
``no slip'' (${\bv} = 0$) condition on the entire boundary the
situation is straightforward, in that it is possible to use the
standard Poincare-Friedrichs inequality: there exists a constant $C_1
> 0$ such that
\[
\|\bv\|_{L^2} \leq C_1 \|\nabla \bv\|_{L^2}\ \ \mbox{for all}\ \bv \in
[H^1_0(\Omega)]^n.
\]  
The use of the classical Poincar\'{e}--Friedrichs inequality is
otherwise identified as a major limitation, even in the conventional
Navier--Stokes related analyses. The Poincar\'{e}--Friedrichs
inequality is only applicable in very limited instances where the
value for the entire boundary is stipulated to be identically zero.
For boundary conditions of a more general nature, such as those
encountered in this study, in which parts of the boundary may be
either a free surface, have an imposed velocity or be subject to
traction conditions, a more suitable inequality is required. (It
should, however, be noted that subtracting a boundary velocity and
analysing the resulting equation is nonetheless still a feasable
alternative, despite the fact that the equations are nonlinear. This
approach requires a more sophisticated and involved level of
mathematics\footnotemark[1].) \footnotetext[1]{{\sc Temam}
\cite{Temam:2} succeeds in arriving at an estimate which proves the
existance of a maximal attractor in two dimensions in this manner.} 

Some common boundary types and associated descriptions are briefly summarised as follows:
\begin{enumerate}
\item {\bf Fixed impermeable boundaries:} The description at such
boundaries is usually Eulerian and the quantities $\tilde {\bF}$ and
${\tilde {\bv}}^{\mbox{\scriptsize {\em ref}}}$  consequently become
identity and zero respectively. These are typically (but not always)
``no slip'' boundaries, implying that ${\tilde
{\bv}}\mid_{\Gamma}={\bf0}$.

\item {\bf Free boundaries:} Conventional use allows the spatial mesh
to slide along free boundaries while still maintaining their overall
Lagrangian character. Stated more formally,
\[
{\tilde {\bn}} \cdot \left( {\tilde {\bv}} - {\tilde {\bv}}^{\mbox{\scriptsize {\em ref}}} \right)=0.
\]
The total volume is nonetheless still a material volume overall.

\item {\bf Imposed velocity-type boundaries:} Conventional use entails descriptions which usually become pure Eulerian at such boundaries. The total flow across such boundaries is zero for an incompressible fluid if volume is to be preserved. For boundary--driven flows one therefore usually assumes that the quantity
\[
\displaystyle \int_{\Gamma} {\bn} \cdot {\bv} \ d \Gamma
\]
vanishes (\'{a} la {\sc Temam} \cite{Temam:2}). 

\item {\bf Imposed traction-type boundaries:} A variety of
descriptions are used at such boundaries, ranging from pure Eulerian
to the vanishing ${\tilde {\bn}} \cdot \left( {\tilde {\bv}} -
{\tilde {\bv}}^{\mbox{\scriptsize {\em ref}}} \right)$ type described
for free boundaries.
\end{enumerate}
These are the modes of reference deformation commonly used at boundaries encountered in practice and which will need to be accommodated if the theory is to be applicable. 

The particular types of geometry considered are those that arise in
problems involving the motion of rigid bodies such as pebbles on the
sea bed; thus a free surface is present, and the domain may be
multiply connected. The Poincar\'{e}--Friedrichs inequality does,
furthermore, not hold on subdomains of the domain in question and the
constant is not optimal.

Further investigation ({\sc communication} \cite{comreddy:1}) reveals
a similar relation, the so-called Poincar\'{e}--Morrey inequality,
holds providing the function attains a value of zero somewhere on the
boundary. The proof of the Poincar\'e-Morrey inequality is, however,
similar to that of one of Korn's inequalities (see, for example, {\sc
Kikuchi} and {\sc Oden} \cite{kikuchi:1}). In particluar, it is
non--constructive, by contradiction and the constant cannot therefore
be determined as part of the proof. Viewed in this light the
forthcoming inequality amounts to a specification of the hypothetical
constant in the Poincar\'{e}--Morrey inequality for domains having a
star--shaped geometry. 
   
\begin{inequality}[A New ``Poincar\'{e}'' Inequality] \label{94}
Suppose ${{\bv}}$ is continuous and differentiable to first order and
that ${{\bv}}$ attains a maximum absolute value, $c$, on an included,
finite neighbourhood of minimum radius $R_{\mbox{\it \scriptsize min}}$
about a point ${{\bx}}^{\mbox{\scriptsize origin}}$ (as depicted in
Fig. \ref{194}).
\begin{figure}[H]
\begin{center} \leavevmode
\mbox{\epsfbox{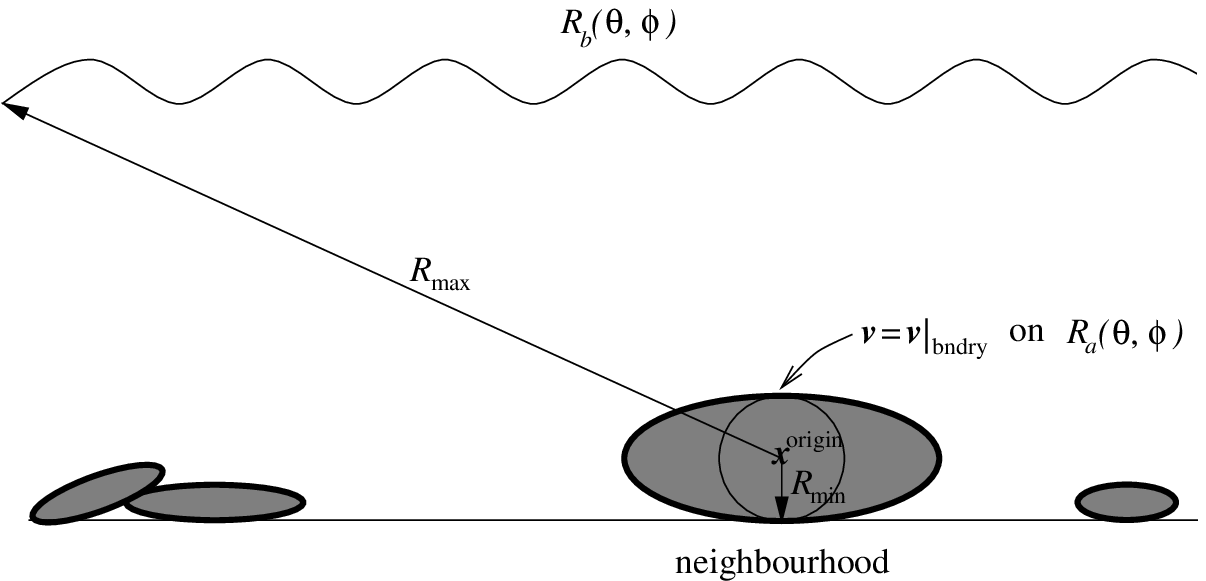}}
%90%
\end{center}
\caption{A Finite Neighbourhood of Minimum Radius $R_{\mbox{\it
\scriptsize min}}$ About a Point ${{\bx}}^{\mbox{\scriptsize
origin}}$.} \label{194}
\end{figure} 
If ${\Omega}$ is a bounded, star--shaped (about a point
${{\bx}}^{\mbox{\scriptsize origin}}$)\footnotemark[1] domain in $R^3$,
then \footnotetext[1]{by which is meant that every point in the domain
can be reached by a straight line from ${{\bx}}^{\mbox{\scriptsize
origin}}$ that does not pass outside of $\Omega$}
\[
\left|\left| {\bv} \right|\right|_{L^2({ \Omega})} \le { \left[
\frac{(R_{\mbox{\it \scriptsize max}} - R_{\mbox{\it \scriptsize min}})
(R_{\mbox{\it \scriptsize max}}^3 - R_{\mbox{\it \scriptsize min}}^3)}
{3 R_{\mbox{\it \scriptsize max}} R_{\mbox{\it \scriptsize min}}}
\right]^{\frac{1}{2}}} \left|\left| { \nabla} { {\bv}}
\right|\right|_{L^2({ \Omega})} + \left|\left| c
\right|\right|_{L^2({\Omega})}
\]
where $R_{\mbox{\it \scriptsize max}}$ is the distance to the farthest
point in ${ \Omega}$ from ${{\bx}}^{\mbox{\scriptsize origin}}$. (Proof in Appendix II.)
\end{inequality}

{\sc Remark:} Notice that ``no slip'' boundaries which contravene the star--shaped requirement are not of any consequence. This is since additional contributions to inequality terms, arising due to the inclusion of any such intruding domains, can be arbitrarily costructed to have a value of zero (without any loss of generality). If for example, one were to apply the inequality in an investigation of the flow around an aeroplane wing, one might make the convenient choice of the wing interior as the desired neighbourhood.

{\sc Remark:} Notice that ``no slip'' boundaries which contravene the star--shaped requirement are not of any consequence. This is since additional contributions to inequality terms, arising due to the inclusion of any such intruding domains, can be arbitrarily costructed to have a value of zero (without any loss of generality). If for example, one were to apply the inequality in an investigation of the flow around an aeroplane wing, one might make the convenient choice of the wing interior as the desired neighbourhood.

This inequality is similar to the Poincar\'{e}--Friedrichs inequality when
$c=0$, but is extended to a geometrical subclass of domains which have
free and partly non-zero boundaries in addition to being more applicable
to more challenging examples such as the flow around an aeroplane wing. It
has a further advantage in that the coefficient is an order of magnitude
more optimal when used under the ``no slip'' Poincar\'{e}--Friedrichs
condition (under such conditions the domain can always be deconstructed
into a number of subdomains in which $R_{\mbox{\it \scriptsize min}} =
\frac{1}{3}R_{\mbox{\it \scriptsize max}}$). The Poincar\'{e}--Friedrichs
inequality is a special case of the above inequality. The necessary lemma
(below) follows naturally from Inequality \ref{94}.

\begin{lemma}[Deviatoric Stress Term Energy] \label{58}
The kinetic energy satisfies the bound $ \ \displaystyle \frac{C}{\rho}
{\tilde K}({\tilde {\bv}}) \ \le \ \left|\left| {\tilde {\bD}}({\tilde
{\bv}}) {\tilde J}^{\frac{1}{2}} \right|\right|_{L^2({\tilde
\Omega})}^2$, where $C$ is related to the constant in Inequality
\ref{94}, $C > 0$. (Proof in Appendix II.)
\end{lemma}

As the reviewers rightly point out, this is Korn's inequality with a
specified constant limited to star--shaped geometries and a variety of
such inequalities can be found on page 323 of {\sc Marsden} and {\sc
Hughes} \cite{mh:1}. The following lemma will facilitate the
elimination of the convective energy rate in the forthcoming
analysis.

\begin{lemma} [Convective Energy Rate] \label{355} The relation
\begin{eqnarray*}
- {\rho} \left< {\tilde {\bv}}, ({\tilde \nabla} {\tilde {\bv}})
{\tilde {\bF}}^{-1} \left( {\tilde {\bv}} - {\tilde
{\bv}}^{\mbox{\scriptsize {\em ref}}} \right) {\tilde J}
\right>_{L^2({\tilde \Omega})} &=& - \frac{1}{2} \rho \left<
{\tilde{\bv}}, {\tilde {\bv}} \left( {\tilde {\bF}}^{-t} {\tilde
{\bN}} \cdot \left( {\tilde {\bv}} - {\tilde
{\bv}}^{\mbox{\scriptsize {\em ref}}} \right) \right) {\tilde J} 
\right>_{L^2(\tilde \Gamma)} \\
&& - \frac{1}{2} \rho \left< {\tilde
{\bv}}, {\tilde {\bv}} \frac{\partial {\tilde J}}{\partial t}
\right>_{L^2({\tilde \Omega})}
\end{eqnarray*}
holds under the conditions required for equations (\ref{307}) and (\ref{308}) to be a completely general reference description. (Proof in Appendix II.)
\end{lemma}

The above lemma is crucial to the analysis for deforming references in particular. The following lemma will establish that the boundary term vanishes at free boundaries under conditions of conventional usage.

\begin{lemma} [Free Boundary Energy Rate] \label{41} The boundary term
\begin{eqnarray*}
- \frac{1}{2} \rho \left< {\tilde{\bv}}, {\tilde {\bv}} \left(
{\tilde {\bF}}^{-t} {\tilde {\bN}} \cdot \left( {\tilde {\bv}} -
{\tilde {\bv}}^{\mbox{\scriptsize {\em ref}}} \right) \right) {\tilde
J}  \right>_{L^2(\tilde \Gamma)} &&
\end{eqnarray*}
vanishes at free boundaries provided the description there is of a
vanishing ${\tilde {\bn}} \cdot \left( {\tilde {\bv}} - {\tilde
{\bv}}^{\mbox{\scriptsize {\em ref}}} \right)$ type. (Proof in
Appendix II.)
\end{lemma}

This concludes the preliminaries required for the deforming reference
energy analysis. 

\subsection{Exponential Dissipation in the Absence of Forcing}

The issue of whether nonlinear, exponential--type dissipation in the
absence of forcing is a property intrinsic to the deforming reference
description is resolved by the following theorem.

\begin{theorem}[Exponential Dissipation in the Absence of Forcing]
\label{60}
A sufficient condition for the completely general reference
description to inherit nonlinear, exponential type energy dissipation
\[
{\tilde K}({\tilde {\bv}}) \le {\tilde K}({\tilde {\bv}} \mid_{t_0})
\ e^{-2 {\nu} C t} 
\]
in the absence of forcing is that the reference moves in a vanishing
${\tilde {\bn}} \cdot \left( {\tilde {\bv}} - {\tilde
{\bv}}^{\mbox{\scriptsize {\em ref}}} \right)$ fashion at free boundaries and becomes pure Eulerian at boundaries of a fixed, impermeable type (the conventional use).
\end{theorem}
{\sc Proof:} Notice that an expression involving the kinetic energy
can be formulated by substituting ${\tilde {\bv}}$ for ${\tilde
{\bw}}$ in the variational momentum equation (\ref{33}). Then
\begin{eqnarray} \label{74}
{\rho} \left< {\tilde {\bv}}, \frac{\partial {\tilde {\bv}}}{\partial
t} {\tilde J} \right>_{L^2(\tilde \Omega)} &=& \left< {\tilde p}
{\tilde \nabla} {\tilde {\bv}} , {\tilde {\bF}}^{-t} {\tilde J}
\right>_{L^2({\tilde \Omega})} - 2 {\mu} \left< {\tilde {\bD}}({\tilde
{\bv}}), {\tilde {\bD}}({\tilde {\bv}}) {\tilde J} \right>_{L^2(\tilde
\Omega)} \nonumber \\
&& - {\rho} \left< {\tilde {\bv}}, ({\tilde \nabla} {\tilde {\bv}})
{\tilde {\bF}}^{-1} \left( {\tilde {\bv}} - {\tilde
{\bv}}^{\mbox{\scriptsize {\em ref}}} \right) {\tilde J}
\right>_{L^2({\tilde \Omega})} \nonumber \\
&& + {\rho} \left< {\tilde {\bv}}, {\tilde {\bb}} {\tilde J}
\right>_{L^2({\tilde \Omega})} + \left< {\tilde{\bv}}, {\tilde{\bP}}
{\tilde {\bN} } \right>_{L^2({\tilde \Gamma})}.
\end{eqnarray}
The order of integration and differentiation are fully
interchangeable (the volume is still a material volume overall for
the type of time--dependent limits associated with free boundaries).
The term containing the pressure, that is 
\[
\left< {\tilde p} {\tilde \nabla} {\tilde {\bv}} :  {\tilde {\bF}}^{-t}
{\tilde J} \right>_{L^2({\tilde \Omega})},
\]
vanishes as a result of incompressibility (equation (\ref{30})). Equation (\ref{74}) can accordingly be rewritten
\begin{eqnarray*}\label{4} 
\frac{1}{2}{\rho} \left( \frac{d}{dt} \left|\left| {\tilde {\bv}}
{\tilde J}^{\frac{1}{2}}  \right|\right|_{L^2({\tilde \Omega})}^2 -
\left< {\tilde {\bv}}, {\tilde {\bv}} \frac{\partial {\tilde
J}}{\partial t} \right>_{L^2({\tilde \Omega})} \right) &=& {\tilde K}_{{\tilde \Gamma} \mbox{\tiny imposed } \mbox{\tiny ${\tilde {\bv}}$}} -2 {\mu}
\left|\left| {\tilde {\bD}} {\tilde J}^{\frac{1}{2}}
\right|\right|_{L^2({\tilde \Omega})}^2 \nonumber \\
&& - {\rho} \left< {\tilde {\bv}}, ({\tilde \nabla} {\tilde {\bv}})
{\tilde {\bF}}^{-1} \left( {\tilde {\bv}} - {\tilde
{\bv}}^{\mbox{\scriptsize {\em ref}}} \right) {\tilde J}
\right>_{L^2({\tilde \Omega})} \nonumber \\
&& + {\rho} \left< {\tilde {\bv}}, {\tilde {\bb}} {\tilde J}
\right>_{L^2(\tilde \Omega)} + \left< {\tilde {\bv}}, {\tilde
{\bP}}{\tilde {\bN}} \right>_{L^2(\tilde \Gamma)}
\end{eqnarray*}
where ${\tilde K}_{{\tilde \Gamma} \mbox{\tiny imposed } \mbox{\tiny ${\tilde
{\bv}}$}}$ is zero for the present by virtue of the fully
interchangeable orders of integration and differentiation (its meaning
will be made clear in the pages to follow). Using Lemmas \ref{58} and
\ref{355} an expression
\begin{eqnarray} \label{59}
\frac{d {\tilde K}({\tilde {\bv}})}{dt} &\le& -2 {\nu} C {\tilde
K}({\tilde {\bv}}) + {\rho} \left< {\tilde {\bv}}, {\tilde {\bb}}
{\tilde J} \right>_{L^2(\tilde \Omega)} + \left< {\tilde {\bv}},
{\tilde {\bP}}{\tilde {\bN}} \right>_{L^2(\tilde \Gamma)} \nonumber \\
&& - \frac{1}{2} \rho \left< {\tilde{\bv}}, {\tilde {\bv}}
\left( {\tilde {\bF}}^{-t} {\tilde {\bN}} \cdot \left( {\tilde {\bv}}
- {\tilde {\bv}}^{\mbox{\scriptsize {\em ref}}} \right) \right)
{\tilde J}  \right>_{L^2(\tilde \Gamma)} + {\tilde K}_{{\tilde \Gamma} \mbox{\tiny imposed } \mbox{\tiny ${\tilde {\bv}}$}}
\end{eqnarray}
is obtained, where ${\tilde K} = \displaystyle \frac{1}{2}
{\rho} \left|\left| {\tilde {\bv}} {\tilde J}^{\frac{1}{2}} 
\right|\right|_{L^2({\tilde \Omega})}^2$ is the total kinetic energy. 

The term ${\tilde {\bF}}^{-t} {\tilde {\bN}} \cdot ({\tilde {\bv}} -
{\tilde {\bv}}^{\mbox{\scriptsize {\em ref}}})$ vanishes at fixed
impermeable boundaries since both $\mbox{${\tilde {\bF}}^{-t}{\tilde
{\bN}}\cdot{\tilde {\bv}}$}$ and ${\tilde {\bv}}^{\mbox{\scriptsize
{\em ref}}}$ vanish under such circumstances (assuming the description becomes purely Eulerian there). This self--same term also vanishes at free boundaries according to Lemma \ref{41}. Boundaries of an imposed velocity type need not be accounted for as a consequence of the stated ``no forcing'' condition, and so
\begin{eqnarray*} \label{54}
\frac{d {\tilde K}({\tilde {\bv}})}{dt} &\le& -2 {\nu} C {\tilde
K}({\tilde {\bv}}) + {\rho} \left< {\tilde {\bv}}, {\tilde {\bb}}
{\tilde J} \right>_{L^2(\tilde \Omega)} + \left< {\tilde {\bv}},
{\tilde {\bP}}{\tilde {\bN}} \right>_{L^2(\tilde \Gamma)}.
\end{eqnarray*}
This equation has a solution of the form
\[
{\tilde K} \le {\tilde K}({\tilde {\bv}} \mid_{t_0}) \ e^{-2 {\nu} C t}
\]
%K({\tilde {\bv}} \mid_{t_0}) e^{-2 {\mu} \left|\left| {\scriptsize
%{\tilde {\bD}}({\tilde {\bv}})} {\tilde J}^{\frac{1}{2}} 
%\right|\right|_{L^2({\tilde \Omega})}^2 t}.
in the absence of forcing ($\mbox{\it ``no forcing''} \Rightarrow
{\tilde {\bb}} = {\tilde {\bP}}{\tilde {\bN}} = {\bf 0}$).
 
A nonlinear, exponential--type energy dissipation in the absence of
forcing is therefore an intrinsic property of the completely general
referential description. This contractive flow property is also an
intrinsic property of the conventional Navier--Stokes equations.

\subsection{Long--Term Stability under Conditions of Time--Dependent
Loading}

The formulation of suitable load and free surface bounds is necessary
before the issue of long-term stability ($L^2$--stability) under
conditions of time--dependent loading can be resolved. The energy
transfer across boundaries at which there is an imposed velocity is a
further factor which must be taken into account under conditions of
forcing. The following lemma facilitates the formulation of load and
free surface bounds.

\begin{lemma}[Force, Free Surface Bounds] \label{49} The inequality
\begin{eqnarray*}
{\rho} \left< {\tilde {\bv}}, {\tilde {\bb}} {\tilde J}
\right>_{L^2({\tilde \Omega})} + \left< {\tilde {\bv}}, {\tilde {\bP}}
{\tilde {\bN}} \right>_{L^2(\tilde \Gamma)} &\le& \displaystyle
\frac{\nu C}{2} \left( {\rho} \left|\left| {\tilde {\bv}} {\tilde
J}^{\frac{1}{2}} \right|\right|_{L^2({\tilde \Omega})}^2 + \left|\left|
{\tilde {\bv}} \right|\right|_{L^2({\tilde \Gamma})}^2 \right) \\
&& + \frac{1}{2 \nu C} \left( {\rho} \left|\left| {\tilde {\bb}}
{\tilde J}^{\frac{1}{2}} \right|\right|_{L^2(\tilde \Omega)}^2 +
\left|\left| {\tilde {\bP}} {\tilde {\bN}} \right|\right|_{L^2(\tilde
\Gamma)}^2 \right)
\end{eqnarray*}
holds where $\nu C$ is a constant, $\nu C > 0$. (Proof in Appendix II.)
\end{lemma}

The relation immediately below will negate any convection--related contribution to the energy bound at imposed velocity--type boundaries.

\begin{lemma} [Convective Energy Rate at an Imposed Velocity--Type Boundary] \label{42} The relation 
\[
- \frac{1}{2} \rho \left< {\tilde{\bv}}, {\tilde {\bv}} \left(
{\tilde {\bF}}^{-t} {\tilde {\bN}} \cdot \left( {\tilde {\bv}} -
{\tilde {\bv}}^{\mbox{\scriptsize {\em ref}}} \right) \right) {\tilde
J}  \right>_{L^2(\tilde \Gamma)} \le 0
\]
holds for boundaries at which there is an imposed velocity provided there is no nett inflow/outflow across such boundaries and the description there becomes pure Eulerian. (Proof in Appendix II.)
\end{lemma}

This done, the mathematical machinery necessary to the long--term
stability analysis is in place. 

\begin{theorem}[Long--Term Stability] \label{61} 
A sufficient condition for the completely general reference
description to inherit the property of long--term stability
\[
\lim_{t \rightarrow \infty} \sup {\tilde K}({\tilde {\bv}}) \le
\frac{M^2}{2 \nu^2 C^2}
\] 
under conditions of time--dependent loading, where this
time--dependent loading, the speed of the surface and any imposed
boundary velocity is bounded in such a way that
\renewcommand{\thefootnote}{\fnsymbol{footnote}}
\begin{eqnarray*} \label{15}
{\rho} \left|\left| {\tilde {\bb}} {\tilde J}^{\frac{1}{2}}  \right|\right|_{L^2(\tilde \Omega)}^2 + \left|\left|
{\tilde {\bP}}{\tilde {\bN}} \right|\right|_{L^2({\tilde \Gamma})}^2 +
\nu^2 C^2 \left|\left| {\tilde {\bv}} \right|\right|_{L^2({\tilde
\Gamma})}^2 + K^2_{{\tilde \Gamma} \mbox{\tiny imposed } \mbox{\tiny ${\tilde {\bv}}$}} \le M^2, \hspace{10mm} \footnotemark[2]
\end{eqnarray*} 
is that the description is of a vanishing ${\tilde {\bn}} \cdot
\left( {\tilde {\bv}} - {\tilde {\bv}}^{\mbox{\scriptsize {\em ref}}}
\right)$ type at free boundaries, that it becomes purely Eulerian at
boundaries across which there is an imposed velocity or where
boundaries are of a fixed, impermeable type. \footnotetext[2]{The
additional terms ${\tilde K}_{{\tilde \Gamma} \mbox{\tiny imposed }
\mbox{\tiny ${\tilde {\bv}}$}}$ are given in Appendix I. They are
only applicable in instances where there is an imposed velocity at
the boundary. The other two boundary terms are only applicable at boundaries which involve tractions.} 
\renewcommand{\thefootnote}{\arabic{footnote}} 
\end{theorem}
{\sc Proof:} Cognizance must now be taken of a previously
unencountered boundary type; that of a stationary boundary across
which there is an imposed velocity. The limits of the integral on the
left hand side of equation (\ref{74}) are time--dependant under such
circumstances and the volume is no longer a material volume overall.
${\tilde K}_{{\tilde \Gamma} \mbox{\tiny imposed } \mbox{\tiny
${\tilde {\bv}}$}}$ in equation (\ref{59}) is no longer zero. Using
Lemmas \ref{41}, \ref{49} and \ref{42} in equation (\ref{59}), then
applying the above bound,
\begin{eqnarray*}
\frac{d {\tilde K}({\tilde {\bv}})}{dt} + \nu C {\tilde K}({\tilde
{\bv}}) &\le& \frac{M^2}{2 \nu C}, 
\end{eqnarray*}
which, when solved, yields 
\begin{eqnarray*}
{\tilde K}({\tilde {\bv}}) &\le& e^{- \nu C t}{\tilde K}({\tilde {\bv}}
\mid_{t=t_0}) + \left( 1 - e^{- \nu C t} \right) \frac{M^2}{2 \nu^2
C^2}.
\end{eqnarray*}
This in turn implies
\begin{eqnarray*}
\lim_{t \rightarrow \infty} \sup {\tilde K}({\tilde {\bv}}) &\le&
\frac{M^2}{2 \nu^2 C^2}.
\end{eqnarray*}
The preceding analyses lead to natural notions of nonlinear
dissipation in the absence of forcing and long--term stability under
conditions of time--dependent loading for the analytic problem. These
properties are also intrinsic features of real flows and the
Navier--Stokes equations.

\section{The Energetic Implications of the Time Discretisation}
\label{6}

This section is concerned with establishing a class of time
discretisations which inherit the self--same energetic properties
(nonlinear dissipation in the absence of forcing and long--term
stability under conditions of time dependent loading) as the analytic
problem, irrespective of the time increment employed. In this section
a generalised, Euler difference time--stepping scheme for the
completely general reference equation is formulated and the energetic
implications are investigated in a similar manner to that carried out
for the analytic equations in the previous section.

This stability analysis is inspired by the approach of others to
schemes for the conventional Navier--Stokes equations. The
desirability of the attributes identified as key energetic properties
is recognised and they have been used as a benchmark in the analysis
of various of the conventional, Eulerian Navier--Stokes schemes by a
host of authors. Related work on the conventional, Eulerian
Navier--Stokes equations can be found in a variety of references, for
example {\sc Temam} \cite{Temam:2} and {\sc Simo} and {\sc Armero}
\cite{s:1}. 

The analyses presented here are extended, not only in the sense that
they deal with the completely general reference equation, but also in
that non--zero boundaries, so--called free boundaries and
time--dependent loads are able to be taken into account (the former
two as a consequence of the new inequality). The findings of this
work have profound consequences for the implementation of the
deforming reference equations. It is significant that many algorithms
used for long--term simulation do not automatically inherit the
fundamental qualitative features of the dynamics.

\subsubsection*{A Generalised Time--Stepping Scheme}

An expression for a generalised Euler difference time--stepping scheme
can be formulated by introducing an ``intermediate'' velocity
\begin{eqnarray} \label{52}
{\tilde {\bv}}_{n + \alpha} \equiv \alpha \ {\tilde {\bv}}\mid_{t +
\Delta t} + (1 - \alpha) \ {\tilde {\bv}}\mid_{t} \hspace{5mm}
\mbox{for} \hspace{5mm} \alpha \in [0, 1]
\end{eqnarray}
to the variational momentum equation (equation (\ref{33}) on page
\pageref{33}) where ${\tilde {\bv}}\mid_t$ and ${\tilde {\bv}}\mid_{t
+ \Delta t}$ are the solutions at times $t$ and $t + \Delta t$
respectively, $\Delta t$ being the time step. It is in this way that
a generalised time--discrete approximation of the momentum equation,
\begin{eqnarray} \label{70}
&& \frac{{\rho}}{\Delta t} \left< {\tilde {\bw}}, ({\tilde {\bv}}_{n +
1} - {\tilde {\bv}}_{n }) {\tilde J}_{n + \alpha}
\right>_{L^2({\tilde \Omega}_{n + \alpha})} = \nonumber \\
&& \hspace{20mm} \left< {\tilde p} {\tilde \nabla} {\tilde {\bw}},
{\tilde {\bF}}_{n + \alpha}^{-t} {\tilde J}_{n + \alpha}
\right>_{L^2({\tilde \Omega}_{n + \alpha})} - 2 \mu \left< {\tilde
{\bD}}({\tilde {\bw}}), {\tilde {\bD}}({\tilde {\bv}}_{n + \alpha})
{\tilde J}_{n + \alpha} \right>_{L^2({\tilde \Omega}_{n +
\alpha})} \nonumber \\
&& \hspace{20mm} - {\rho} \left<{\tilde {\bw}}, ({\tilde \nabla}
{\tilde {\bv}}_{n + \alpha}) {\tilde {\bF}}_{n + \alpha}^{-1} \left(
{\tilde {\bv}}_{n + \alpha} - {\tilde {\bv}}_{n +
\alpha}^{\mbox{\scriptsize {\em ref}}} \right) {\tilde J}_{n + \alpha}
\right>_{L^2({\tilde \Omega}_{n + \alpha})} \nonumber \\
&& \hspace{20mm} + {\rho} \left<{\tilde {\bw}}, {\tilde {\bb}}_{n +
\alpha} {\tilde J}_{n + \alpha} \right>_{L^2({\tilde \Omega}_{n +
\alpha})} + \left< {\tilde{\bw}} , {\tilde{\bP}}_{n + \alpha} {\tilde
{\bN}_{n + \alpha}} \right>_{L^2({\tilde \Gamma}_{n + \alpha})},
\end{eqnarray}
is derived, where $\left< \ . \ \right>_{L^2({\tilde \Omega}_{n +
\alpha})}$ denotes the $L^2$ inner product over the deforming domain at
time $t + \alpha \Delta t$. ${\tilde \Gamma}_{n + \alpha}$, ${\tilde
{\bF}}_{n + \alpha}$, ${\tilde J}_{n + \alpha}$, ${\tilde {\bD}}_{n +
\alpha}$, ${\tilde {\bP}}_{n + \alpha}$, and ${\tilde {\bb}}_{n +
\alpha}$ are likewise defined to be the relevant quantities evaluated
at time $t + \alpha \Delta t$.

It will presently become apparent that relevant energy terms are not
readily recovered from the time-discrete equations for deforming
references in general. It may therefore make sense to perform the
analyses for the time--discrete equation in the context of divergence
free rates of reference deformation only. A practically less
restrictive alternative is too labour intensive. This investigation
is accordingly restricted to a subclass of reference deformations in
which ``reference volume'' is conserved. This is for reasons of
expedience alone and it is hoped that this subclass of deformations
is thought to be representative.

\begin{assumption} \label{assumption} The assumptions ${\tilde J}_{n}
= {\tilde J}_{n + \alpha}$ and ${\tilde J}_{n + 1} = {\tilde J}_{n +
\alpha}$ are made so that the desired energy terms are readily recovered as
\[
{\tilde K}({\tilde {\bv}}_n) = \frac{1}{2} \rho \left|\left| {\tilde
{\bv}}_n {\tilde J}_n^{\frac{1}{2}} \right|\right|_{L^2({\tilde
\Omega}_n)}^2 = \frac{1}{2} \rho \left|\left| {\tilde {\bv}}_{n}
{\tilde J}_{n + \alpha}^{\frac{1}{2}} \right|\right|_{L^2({\tilde
\Omega}_{n + \alpha})}^2 
\]
and
\[
{\tilde K}({\tilde {\bv}}_{n+1}) = \frac{1}{2} \rho \left|\left|
{\tilde {\bv}}_{n+1} {\tilde J}_{n+1}^{\frac{1}{2}}
\right|\right|_{L^2({\tilde \Omega}_{n+1})}^2 = \frac{1}{2} \rho
\left|\left| {\tilde {\bv}}_{n+1} {\tilde J}_{n +
\alpha}^{\frac{1}{2}} \right|\right|_{L^2({\tilde \Omega}_{n +
\alpha})}^2.
\]
\end{assumption} 

{\sc Remark:} Notice that $\displaystyle \frac{{\tilde J}_{n + 1} -
{\tilde J}_{n}}{\Delta t} = {\tilde J}_{n + \alpha} \mathop{\rm
div}{\bv}^{\mbox{\scriptsize {\it ref}}}_{n + \alpha}$, the discrete
form of $\displaystyle \displaystyle \frac{\partial {\tilde
J}}{\partial t} = {\tilde J} \mathop{\rm div}{\bv}^{\mbox{\scriptsize
{\it ref}}}$, can consequently be rewritten as 
\[
\mathop{\rm
div}{\bv}^{\mbox{\scriptsize {\it ref}}}_{n + \alpha} = 0
\]
under the conditions of the above assumption. It is for the practical
expedience afforded by Assumption \ref{assumption} alone that this
analysis is limited to instances in which $\mathop{\rm
div}{\bv}^{\mbox{\scriptsize {\it ref}}}_{n + \alpha} = 0$.

The following lemma will facilitate the elimination of the rate of
energy change associated with the convective term under these
conditions.

\begin{lemma} [Discrete Convective Energy Rate] \label{354}
The following relation involving the discrete convective term holds for an incompressible fluid under circumstances of $\mathop{\rm
div}{\bv}^{\mbox{\scriptsize {\it ref}}}_{n + \alpha} = 0$:
\begin{eqnarray*}
- {\rho} \left<{\tilde {\bv}}_{n + \alpha}, ({\tilde \nabla}
{\tilde {\bv}}_{n + \alpha}) {\tilde {\bF}}_{n + \alpha}^{-1} \left(
{\tilde {\bv}}_{n + \alpha} - {\tilde {\bv}}_{n +
\alpha}^{\mbox{\scriptsize {\em ref}}} \right) {\tilde J}_{n +
\alpha} \right>_{L^2({\tilde \Omega}_{n + \alpha})} \hspace{100mm} && \\
\hspace{50mm} = - \frac{1}{2} \rho \left< {\tilde {\bv}}_{n +
\alpha}, {\tilde {\bv}}_{n + \alpha} \left( {\tilde {\bF}}_{n +
\alpha}^{-t} {\tilde {\bN}} \cdot ({\tilde {\bv}}_{n + \alpha} -
{\tilde {\bv}}_{n + \alpha}^{ref}) \right) {\tilde J}_{n + \alpha} 
\right>_{L^2(\tilde \Gamma_{n + \alpha})}. \hspace{40mm} &&             \end{eqnarray*}
(Proof in Appendix II.)
\end{lemma}

{\sc Remark:}\label{assumpremark} Recall that in the investigation of
the analytic problem, a term arising from the manipulation of the
acceleration containing term (the term containing the rate of change
of the Jacobian) cancelled with the convective energy. It is
therefore not surprising that assumptions pertaining to the
acceleration containing term (in particular to the rate of change of
the Jacobian) in the discrete problem will, once made, also be
necessary for the corresponding discrete convective energy term to
vanish (reffering to the $\mathop{\rm div}{\bv}^{\mbox{\scriptsize
{\it ref}}} = 0$ condition of Lemma \ref{354}). This is a good
prognosis for the energetic behaviour of the discrete problem in
circumstances of reference deformations excluded by Assumption
\ref{assumption}. The full ramifications of Assumption
\ref{assumption} are considered in Subsection \ref{45} of Appendix I.
This concludes the preliminaries required for the analysis of the
time--discrete equation.

\subsection{Nonlinear Dissipation in the Absence of Forcing}

The following analysis establishes a class of time--stepping schemes
which exhibit nonlinear dissipation in the absence of forcing
regardless of the time increment employed.

\begin{theorem}[Nonlinear Dissipation in the Absence of Forcing]
\label{55} Suppose that the description is of a vanishing ${\tilde
{\bn}}_{n+\alpha} \cdot \left( {\tilde {\bv}}_{n+\alpha} - {\tilde
{\bv}}^{\mbox{\scriptsize {\em ref}}}_{n+\alpha} \right)$ type at
free boundaries, purely Eulerian at boundaries across which there is
an imposed velocity and that the deformation rate of the reference is
divergence free. A sufficient condition for the kinetic energy
associated with the generalised class of time--stepping schemes to
decay nonlinearly
\begin{eqnarray*}
{\tilde K}({\tilde {\bv}}_{n + 1}) - {\tilde K}({\tilde {\bv}}_{n })
&\le& - {\Delta t} \ 2 \mu \left|\left| {\tilde {\bD}}({\tilde
{\bv}}_{n + \alpha}) {\tilde J}_{n + \alpha}^{\frac{1}{2}} 
\right|\right|_{L^2({\tilde \Omega}_{n + \alpha})}^2
\end{eqnarray*}
in the absence of forcing and irrespective of the time increment
employed, is that the scheme is as, or more, implicit than central
difference. That is
\[
\alpha \ge \frac{1}{2}.
\] 
\end{theorem}
{\sc Proof:} Expressing the ``intermediate'' velocities ${\tilde
{\bv}}_{n + \frac{1}{2}}$ and ${\tilde {\bv}}_{n + \alpha}$ in terms
of equation (\ref{52}) and subtracting, the result
\begin{eqnarray} \label{48}
{\tilde {\bv}}_{n + \alpha} = \left( \alpha - \frac{1}{2} \right)
\left( {\tilde {\bv}}_{n  + 1} - {\tilde {\bv}}_{n } \right) +
{\tilde {\bv}}_{n + \frac{1}{2}}
\end{eqnarray}
is obtained. The first step towards formulating an expression
involving the kinetic energy of the generalised time stepping--scheme
(\ref{70}) is to replace the arbitrary vector, ${\bw}$, with ${\tilde
{\bv}}_{n + \alpha}$. By further substituting (\ref{48}) into
(\ref{70}) and eliminating the pressure containing term on the basis of incompressibility (equation (\ref{30})), an expression involving the
difference in kinetic energy over the duration of a single time step
is obtained.

Incompressibility and a restriction on reference deformations to
those for which ${\mathop {\rm div}}{\bv}^{\scriptsize
ref}_{n+\alpha}$ is zero ensure that the Lemma \ref{354} condition is
satisfied.

The equation 
\begin{eqnarray} \label{50}
{\tilde K}({\tilde {\bv}}_{n  + 1}) - {\tilde K}({\tilde {\bv}}_{n })
&=& - {\rho} \left( \alpha - \frac{1}{2} \right) \left|\left| \left(
{\tilde {\bv}}_{n  + 1} - {\tilde {\bv}}_{n} \right) {\tilde J}_{n +
\alpha}^{\frac{1}{2}} \right|\right|_{L^2({\tilde \Omega}_{n +
\alpha})}^2 \nonumber \\
&& - {\Delta t} \ 2 \mu \left|\left| {\tilde {\bD}}({\tilde {\bv}}_{n
+ \alpha}) {\tilde J}_{n + \alpha}^{\frac{1}{2}} 
\right|\right|_{L^2({\tilde \Omega}_{n + \alpha})}^2 \nonumber +
{\Delta t} {\rho} \left< {\tilde {\bv}}_{n + \alpha}, {\tilde
{\bb}}_{n + \alpha} {\tilde J}_{n + \alpha} \right>_{L^2({\tilde
\Omega}_{n + \alpha})} \nonumber \\
&& + {\Delta t} \left< {\tilde {\bv}}_{n + \alpha}, {\tilde{\bP}}_{n
+ \alpha} {\tilde{\bN}}_{n + \alpha} \right>_{L^2({\tilde \Gamma}_{n
+ \alpha})} \nonumber \\
&& - {\Delta t}\frac{1}{2}\rho \left< {\tilde {\bv}}_{n +
\alpha}, {\tilde {\bv}}_{n + \alpha} \left( {\tilde {\bF}}_{n +
\alpha}^{-t} {\tilde {\bN}} \cdot ({\tilde {\bv}}_{n + \alpha} -
{\tilde {\bv}}_{n + \alpha}^{ref}) \right) {\tilde J}_{n + \alpha} 
\right>_{L^2(\tilde \Gamma_{n + \alpha})},
\end{eqnarray}
is then obtained. The term ${\tilde {\bF}}^{-t}_{n+\alpha} {\tilde
{\bN}}_{n+\alpha} \cdot ({\tilde {\bv}}_{n+\alpha} - {\tilde
{\bv}}_{n+\alpha}^{\mbox{\scriptsize {\em ref}}})$ vanishes at fixed
impermeable boundaries since both $\mbox{${\tilde
{\bF}}^{-t}_{n+\alpha}{\tilde {\bN}}_{n+\alpha}\cdot{\tilde
{\bv}}_{n+\alpha}$}$ and ${\tilde {\bv}}^{\mbox{\scriptsize {\em
ref}}}_{n+\alpha}$ vanish under such circumstances (assuming the
description becomes purely Eulerian there). This self--same term also
vanishes at free boundaries according to Lemma \ref{41}. Boundaries
of an imposed velocity type need not be accounted for as a
consequence of the stated ``no forcing'' condition, and so
\begin{eqnarray*}
{\tilde K}({\tilde {\bv}}_{n + 1}) - {\tilde K}({\tilde {\bv}}_{n })
&\le& - {\rho} \left( \alpha - \frac{1}{2} \right) \left|\left| \left(
{\tilde {\bv}}_{n + 1} - {\tilde {\bv}}_{n} \right) {\tilde J}_{n +
\alpha}^{\frac{1}{2}} \right|\right|_{L^2({\tilde \Omega}_{n +
\alpha})}^2 \\
&& - {\Delta t} \ 2 \mu \left|\left| {\tilde {\bD}}({\tilde {\bv}}_{n +
\alpha}) {\tilde J}_{n + \alpha}^{\frac{1}{2}}
\right|\right|_{L^2({\tilde \Omega}_{n + \alpha})}^2,
\end{eqnarray*} 
because of this condition. Thus the kinetic energy inherent to the
algorithmic flow decreases nonlinearly in the absence of forcing,
irrespective of the time increment employed and for arbitrary initial
conditions provided that
\begin{eqnarray*}
\alpha \ge \frac{1}{2} \hspace{10mm} \mbox{and} \hspace{10mm}
\mathop{\rm div} {\bv}_{n + \alpha}^{\mbox{\scriptsize {\em ref}}} =
0.
\end{eqnarray*}
The former requirement translates directly into one specifying the
use of schemes as, or more, implicit than central difference. Only
for descriptions which are divergence free has it here been guaranteed
that energy will not be artificially introduced by way of the
reference.

{\sc Remark:} Notice (by Lemma \ref{58}) that for $\alpha =
\frac{1}{2}$ an identical rate of energy decay
\begin{eqnarray*}
\frac{{\tilde K}({\tilde {\bv}}_{n + 1}) - {\tilde K}({\tilde {\bv}}_{n
})}{{\Delta t}} &\le& - 2 \nu C {\tilde K}({\tilde {\bv}}_{n + \alpha})
\end{eqnarray*} 
is obtained for the discrete approximation as was obtained for the
equations.

\subsection{Long--Term Stability under Conditions of Time--Dependent
Loading}

This second part of the time--discrete analysis establishes a class
of time stepping schemes which exhibit long--term stability under
conditions of time dependent loading, irrespective of the time
increment employed. The following lemma is necessary to the analysis
and is concerned with devising a bound for the energy at an
intermediate point in terms of energy values at either end of the
time step.
\begin{lemma}[Intermediate Point Energy] \label{57} The following
bound applies
\[
{\tilde K}({\tilde {\bv}}_{n + \alpha}) \ge \alpha \left( \alpha - c +
\alpha c \right) {\tilde K}({\tilde {\bv}}_{n  + 1}) + (1-\alpha)
\left( 1-\alpha - \displaystyle \frac{\alpha}{c} \right) {\tilde
K}({\tilde {\bv}}_{n })
\]
where $c$ is some constant, $c > 0$.
\end{lemma}

The optimal choice of the constant $c$ is established farther on. The
following theorem establishes a class of time--stepping schemes which
exhibit long--term stability under conditions of time--dependent
loading regardless of the time increment employed.

\begin{theorem}[Long--Term Stability] \label{155}
Suppose that the description is of a vanishing ${\tilde
{\bn}}_{n+\alpha} \cdot \left( {\tilde {\bv}}_{n+\alpha} - {\tilde
{\bv}}^{\mbox{\scriptsize {\em ref}}}_{n+\alpha} \right)$ type at
free boundaries, that it becomes purely Eulerian at fixed impermeable
boundaries and that the rate at which the reference is deformed is
divergence free. A sufficient condition for the algorithmic flow to
exhibit long--term stability under conditions of time--dependent
loading assuming this time--dependent loading and the speed of the
free surface is bounded in such a way that
\renewcommand{\thefootnote}{\fnsymbol{footnote}}
\begin{eqnarray*}
{\rho} \left|\left| {\tilde {\bb}}_{n + \alpha} {\tilde J}_{n +
\alpha}^{\frac{1}{2}} \right|\right|_{L^2({\tilde \Omega}_{n +
\alpha})}^2 + \left|\left| {\tilde {\bP}}_{n + \alpha}{\tilde
{\bN}}_{n + \alpha} \right|\right|_{L^2({\tilde \Gamma}_{n +
\alpha})}^2 + \nu^2 C^2 \left|\left| {\tilde {\bv}}_{n + \alpha}
\right|\right|_{L^2({\tilde \Gamma}_{n + \alpha})}^2 &\le& M^2,
\hspace{5mm} \footnotemark[2]
\end{eqnarray*} \footnotetext[2]{Note that this bound does not incorporate a contribution from boundaries of an imposed velocity type in any obvious way. The two boundary terms are only applicable at boundaries which involve tractions.} 
\renewcommand{\thefootnote}{\arabic{footnote}}  
is
\[
\alpha > \frac{1}{2}.
\] 
\end{theorem}
{\sc Proof:} Substituting Lemmas \ref{58}, \ref{41} and \ref{49} into equation (\ref{50}), applying the
above bound and choosing $\alpha \ge \frac{1}{2}$ one obtains
\begin{eqnarray*}
\frac{{\tilde K}({\tilde {\bv}}_{n  + 1}) - {\tilde K}({\tilde
{\bv}}_{n })}{\Delta t} + \nu C {\tilde K}({\tilde {\bv}}_{n + \alpha})
&\le& \frac{M^2}{2 \nu C}.
\end{eqnarray*}
From this point on the argument used is identical to that of {\sc
Simo} and {\sc Armero} \cite{s:1} for the conventional, Eulerian
Navier--Stokes equations.  Substitution of Lemma \ref{57} leads to a
recurrence relation,
\begin{eqnarray*}
{\tilde K}({\tilde {\bv}}_{n + 1}) &\le& \frac{1 - \nu C (1 - \alpha)(1
- \alpha - \frac{\alpha}{c})\Delta t}{1 + \nu C\alpha (\alpha - c +
\alpha c)\Delta t}{\tilde K}({\tilde {\bv}}_{n }) + \frac{M^2 \Delta
t}{2 \nu C \left[ 1 + \nu C\alpha (\alpha - 1 + \alpha c)\Delta t
\right]}.
\end{eqnarray*}
Using this recurrence relation to take cognisance of the energy over
all time steps,
\begin{eqnarray} \label{200}
{\tilde K}({\tilde {\bv}}_{n + 1}) &\le& \left[ \frac{1 - \nu C(1 -
\alpha)(1 - \alpha - \frac{\alpha}{c})\Delta t}{1 + \nu C\alpha (\alpha
- c + \alpha c)\Delta t} \right]^{n} {\tilde K}({\tilde {\bv}}_0)
\nonumber \\
&& + \frac{M^2 \Delta t}{2\nu C \left[ 1 + \nu C\alpha (\alpha - c +
\alpha c)\Delta t \right]} \sum_{k=0}^{n-1} \left[ \frac{(1 - \nu C(1 -
\alpha)(1 - \alpha - \frac{\alpha}{c})\Delta t)}{1 + \nu C\alpha
(\alpha - c + \alpha c)\Delta t} \right]^{k} \nonumber \\
&& \hspace{133mm}
\end{eqnarray}
is obtained. An infinite geometric series which converges so that
\begin{eqnarray*}
\lim_{n \rightarrow \infty} \sup {\tilde K}({\tilde {\bv}}_{n  + 1})
&\le& \frac{M^2 \Delta t}{2 \nu C \left[ 1 + \nu C \alpha (\alpha - c +
\alpha c)\Delta t \right]} \left[ 1 \frac{}{}^{}_{} \right. \\
&& \hspace{30mm} \left. - \ \frac{(1 - \nu C(1 - \alpha)(1 - \alpha -
\frac{\alpha}{c})\Delta t)}{1 + \nu C \alpha (\alpha - c + \alpha
c) \Delta t} \right]^{-1} \\
&& \\
&=& \frac{M^2}{2\nu C \left[ \nu C\alpha (\alpha - c + \alpha c) + \nu
C(1 - \alpha)(1 - \alpha - \frac{\alpha}{c}) \right]}
\end{eqnarray*}
results, providing the absolute ratio of the series is less than
unity. That is
\[
\left| \frac{1 - \nu C(1 - \alpha)(1 - \alpha - \frac{\alpha}{c})\Delta
t}{1 + \nu C\alpha (\alpha - c + \alpha c)\Delta t} \right| < 1.
\]
Therefore either
\[
- 1 - \nu C\alpha(\alpha - c + \alpha c)\Delta t < 1 - \nu C(1 -
\alpha) \left (1 - \alpha - \frac{\alpha}{c} \right) \Delta t
\]
or
\begin{eqnarray} \label{202}
1 - \nu C(1 - \alpha) \left (1 - \alpha - \frac{\alpha}{c} \right)
\Delta t < 1 + \nu C\alpha(\alpha - c + \alpha c)\Delta t
\end{eqnarray}
in order for the bound to exist. Notice, furthermore, that for this
desired convergence to be unconditional (regardless of the time
increment employed) requires
\begin{eqnarray} \label{201}
\alpha - c + \alpha c \ge 0.
\end{eqnarray} 
The denominator in the series ratio might otherwise vanish for some
value of $\Delta t$.

For $\alpha \in \left[\frac{1}{2}, 1\right]$ equation (\ref{202}) and
equation (\ref{201}) together imply
\begin{eqnarray*}
\frac{(1 - \alpha)}{\alpha} < c \le \frac{\alpha}{(1 -
\alpha)} 
\end{eqnarray*}
which in its turn implies
\begin{eqnarray*}
\frac{(1 - \alpha)}{\alpha} < \frac{\alpha}{(1 - \alpha)}.
\end{eqnarray*}
The choice of the parameter $\alpha > \frac{1}{2}$ therefore leads to
an infinite geometric series which forms the desired upper bound. The
minimum value of this bound occurs for $c$ chosen according to
\begin{eqnarray*}
\inf_{\frac{(1-\alpha)}{\alpha} < c \le \frac{\alpha}{(1-\alpha)}}
\frac{1}{\nu C\alpha(\alpha - c + \alpha c) + \nu C(1 - \alpha)(1 -
\alpha - \frac{\alpha}{c})} &=& \frac{1}{\nu C (2\alpha - 1)^2}.
\end{eqnarray*}
The value of this upper bound, which occurs for the choice of the
parameter $\alpha > \frac{1}{2}$, is then
\begin{eqnarray*}
\lim_{n \rightarrow \infty} \sup {\tilde K}({\tilde {\bv}}_{n  + 1})
&\le& \frac{M^2}{2 \nu^2 C^2 (2\alpha - 1)^2}.
\end{eqnarray*}
In this way one arrives at a class of algorithms which are
unconditionally (irrespective of the time increment employed) stable.

{\sc Remark:} Notice that for $\alpha = 1$ one obtains an identical
energy bound for the discrete approximation as was obtained for the
equations.

\section{Some Numerical Examples}

Some numerical results for problems of the type in question are
presently given. The theory thus far developed was employed in the
simulation of a driven cavity flow, a driven cavity flow with various,
included rigid bodies, a die--swell problem and a Stokes,
second order wave. 

The approach taken when approximating free surfaces, was
that they may be treated as a material entity, that is, the material
derivative of the free surface was assumed zero. Euler's equations and
conservation of linear momentum were used to determine the motion of
the rigid body. A predictor--corrector method was used to solve the
combined sub--problems. 

A backward difference scheme was used to approximate the time
derivative in the fluid sub--problem (in compliance with the Theorem
\ref{55} and Theorem \ref{155} conditions), the finite element method
was used for the spatial (referential ``space'') discretisation and a
$Q_2$--$P_1$ element pair was used as a basis. A penalty method was employed to eliminate pressure as a variable and nonlinearity was
circumvented by way of a new, second order accurate linearisation.
Linearising with a guess obtained by extrapolating through solutions
from the previous two time steps leads to second order accuracy. 
\begin{theorem} \label{148}
The linearised terms, $ ( 2 {\bv} \mid_t - {\bv} \mid_{t-\Delta t} )
\cdot \nabla {\bv} \mid_{t+\Delta t} $ and $ {\bv} \mid_{t+\Delta t}
\cdot \nabla (2 {\bv} \mid_t - {\bv} \mid_{t-\Delta t})$, are second
order accurate (have error $ O(\Delta t^2)$) approximations of the
nonlinear term $({\bv} \cdot \nabla {\bv}) \mid_{t+\Delta t}$.
\end{theorem}
{\sc Proof:} 
\begin{eqnarray*}
{\bv} \mid_{t+\Delta t} &=& {\bv} \mid_t + \Delta t \left.
\frac{\partial {\bv}}{\partial t} \right|_{t} + O(\Delta t^2)
\hspace{10mm} \mbox{\it (by Taylor series)} \\
&=& {\bv} \mid_{t} + \Delta t \left( \frac{ {\bv} \mid_t - {\bv}
\mid_{t-\Delta t} }{ \Delta t } + O(\Delta t) \right) + O(\Delta t^2)
\hspace{10mm}  \mbox{\it (using a backward} \\
& & \hspace{90mm} \mbox{\it difference)} \\
&=& 2{\bv}\mid_t - {\bv}\mid_{t-\Delta t} + O(\Delta t^2) \\
\left( {\bv} \cdot \nabla {\bv} \right) \mid_{t+\Delta t} &=& \left[
2{\bv}\mid_t - {\bv}\mid_{t-\Delta t} + O(\Delta t^2) \right]
\frac{}{} \cdot {\nabla} {\bv} \mid_{t+\Delta t} \\
&=& \left[ 2{\bv}\mid_t - {\bv}\mid_{t-\Delta t} \right] \cdot \nabla
{\bv} \mid_{t+\Delta t} + O(\Delta t^2) \frac{}{}
\end{eqnarray*}
\renewcommand{\thefootnote}{\fnsymbol{footnote}}
The above linearisation schemes are an improvement on the
conventional ${\bv}\mid_t \cdot \nabla {\bv} \mid_{t+\Delta t}
\footnotemark[2]$ \footnotetext[2]{Favoured in terms of both rate and
radius of convergence by {\sc Cuvelier}, {\sc Segal} and {\sc van
Steenhoven} \cite{c:3}.} or ${\bv} \mid_{t+\Delta t} \cdot \nabla
{\bv} \mid_t$ linearisation schemes by an order of magnitude. A
detailed exposition of all numerical methods otherwise used in these
simulations can be found in {\sc Childs} and {\sc Reddy}
\cite{me:2}.
\renewcommand{\thefootnote}{\arabic{footnote}}  

\subsection{Example 1: Driven Cavity Flow} \label{32} 

\begin{figure}[H]
\begin{center} \leavevmode
\mbox{\epsfbox{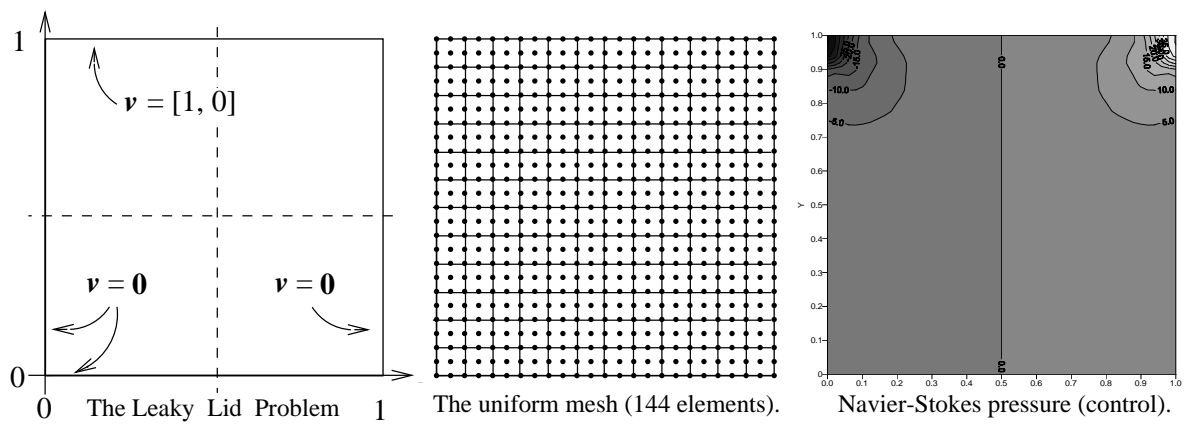}}
\end{center}
\caption{The Problem, the Mesh and the Pressures Obtained Using the
Conventional Eulerian Equations.} \label{82}
\end{figure}

\begin{figure}[H]
\begin{center} \leavevmode
\mbox{\epsfbox{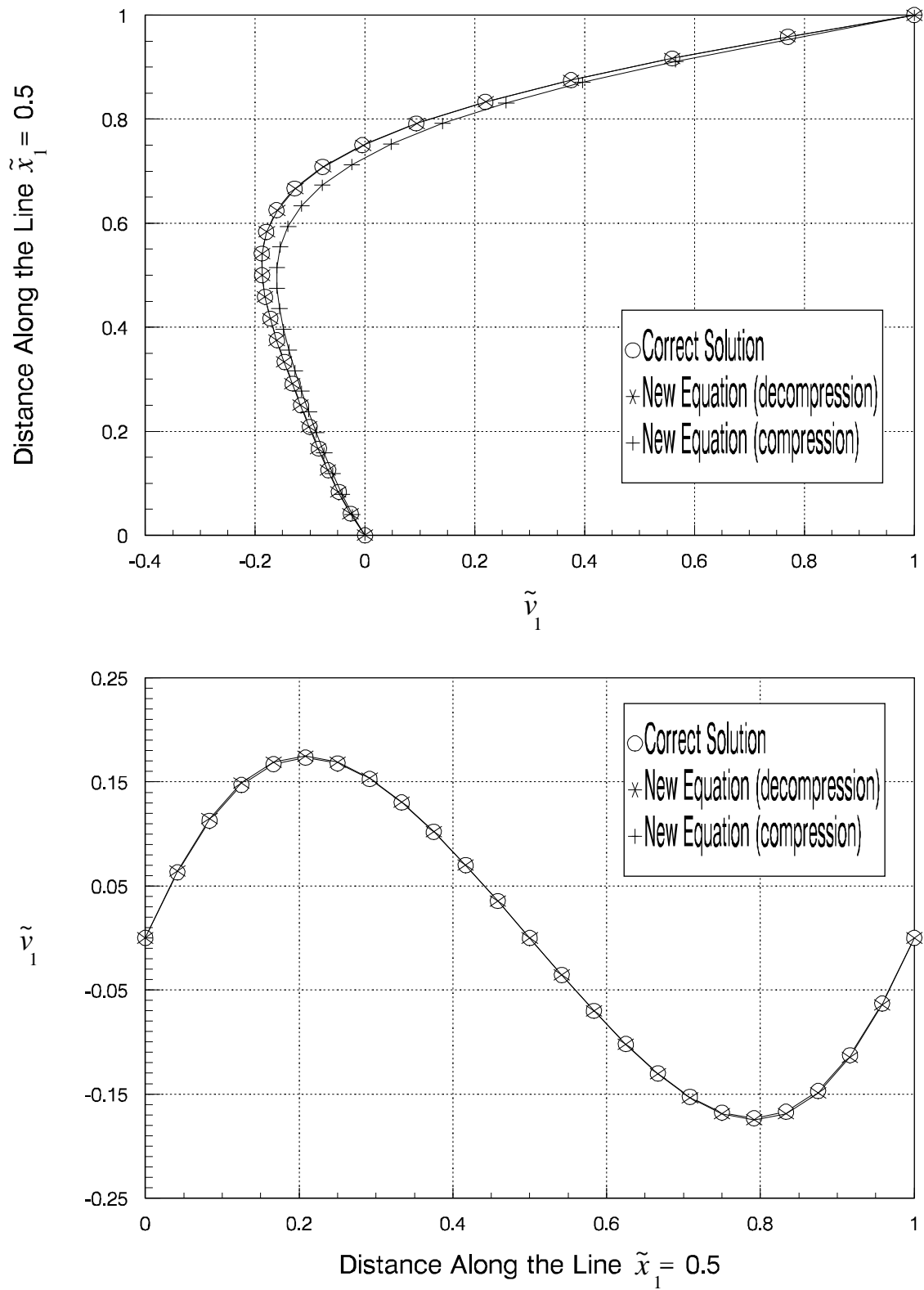}}
\end{center}
\caption{In this test part of the mesh was successively compressed
and decompressed by 5 \% over two time steps of length 0.05.}
\label{84}
\end{figure}

The problem is essentially that of a square, two--dimensional pot
whose lid is moved across the top at a rate equal to its diameter for
a Reynolds number of unity. The boundary conditions are accordingly
``no slip'' on container walls and a horizontal flow of unity across
the top (depicted in Fig. \ref{82}). 
The idea here was to compare results obtained using the completely general reference equation on a deforming mesh with those obtained using the conventional, Eulerian Navier--Stokes equations. 
\begin{figure}[H]
\begin{center} \leavevmode
\mbox{\epsfbox{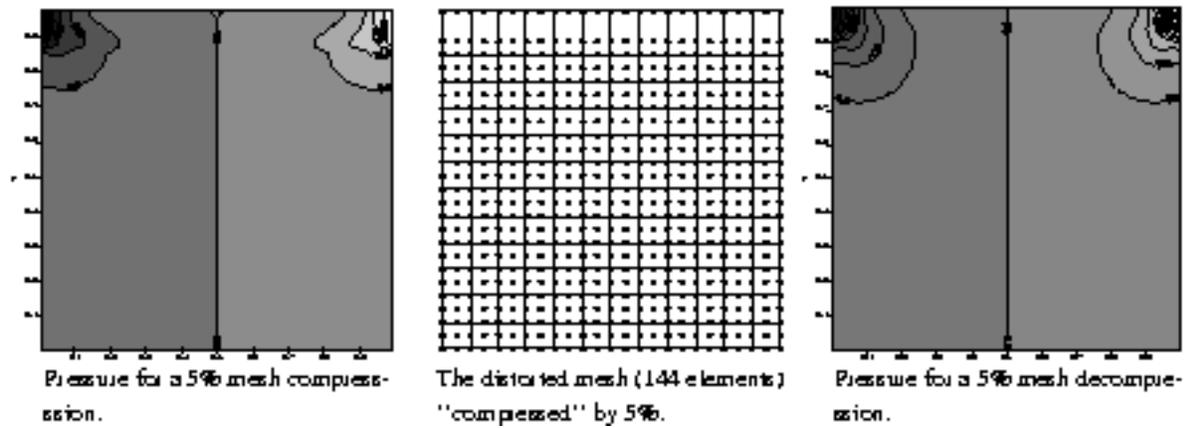}}
\end{center}
\caption{Pressures Obtained Using the Completely General Reference
Equation and a Deforming Mesh.} \label{76}
\end{figure}
The corresponding velocity profiles along the cuts depicted in Fig. \ref{82} are given in Fig. \ref{84}.
\newpage
\begin{center} {\large Velocity Profiles} \end{center}
\vspace{10mm}
\newpage

\subsection{Example 2: ``Pebble in a Pothole''}

In this example rigid bodies of varying mass and moments of inertia
were released from rest in a flow dictated by the same boundary
conditions as the driven cavity flow of the previous example. One
would expect a die bead (a small rigid body of neutral bouyancy)
to move in tandem with the fluid soon after its release from rest.
One might also expect a clockwise rotation to be induced by
concentrating the mass closer to the centre i.e. lowering the moment
of inertia.

The finite element mesh was automatically generated and adjusted
about the included rigid body in what is possibly a slightly novel
fashion. A small region of mesh immediately adjacent to the included
rigid body was repeatedly remapped to cope with the changing
orientation, the remainder was squashed/stretched according to the
translation.

\begin{figure}[H]
\begin{center} \leavevmode
\mbox{\epsfbox{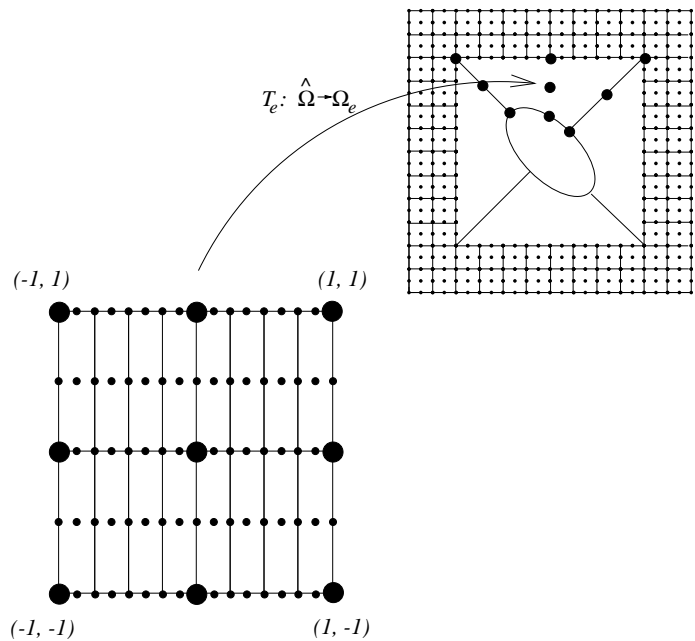}}
%35%
\end{center}
\caption{The Local Distortion is Obtained by Mapping Square Chunks of
Rectangular Mesh Using Finite Element Mappings.} \label{3}
\end{figure}

To begin with, a square region of mesh centered on, and including the
rigid body, is deleted (depicted in Fig. \ref{3}). Each of four
wedge--shaped regions is then demarcated (the intersections of lines
which bisect corners and edges of the square frame, with the surface
of the rigid body are located using Newton's method) by as many
points as there are nodes in an element i.e. each wedge
shaped--region is set up as a massive element.
Chunks of uniform mesh, which have identical extremities to those of
the master element, are then mapped into the newly--demarcated,
wedge--shaped regions using finite element mappings (in exactly the
same manner as points in the master element domain are, in theory,
mapped into individual mesh elements). Further, fine adjustment of
nodes intended to delineate the surface of the rigid body is
accomplished by moving them along a line between node and centre, to
the rigid body surface using Newton's method. The mesh outside the
``box'' (the box containing the 4 wedges enclosing the rigid body) is
squashed/stretched according to the requirements of the translation
(the nodes are translated by a factor inversely proportional to their
distance from the box). This method satisfies the requirement that
${\tilde {\bn}} \cdot \left( {\tilde {\bv}} - {\tilde
{\bv}}^{\mbox{\scriptsize {\em ref}}} \right)$ vanishes at the
fluid--rigid body interface (a condition in Theorems \ref{60},
\ref{61}, \ref{55} and \ref{155}). Mesh refinement in the vicinity of
the included, rigid body is an automatic by-product of this method. 
\begin{figure}[H]
\begin{center} \leavevmode
\mbox{\epsfbox{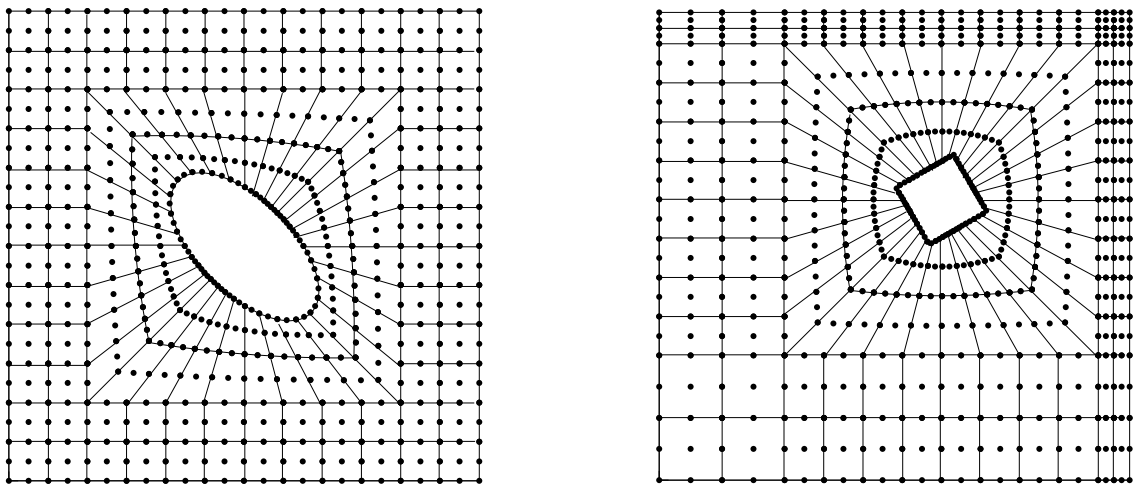}}
%50%
\end{center}
\caption{Typical meshes which result when using this method of automatic mesh generation about rigid bodies which are simultaneously rotating and translating.}
\end{figure}

\begin{figure}[H]
\begin{center} \leavevmode
%120%
%%BoundingBox: 25 20 502 442
\mbox{\epsfbox{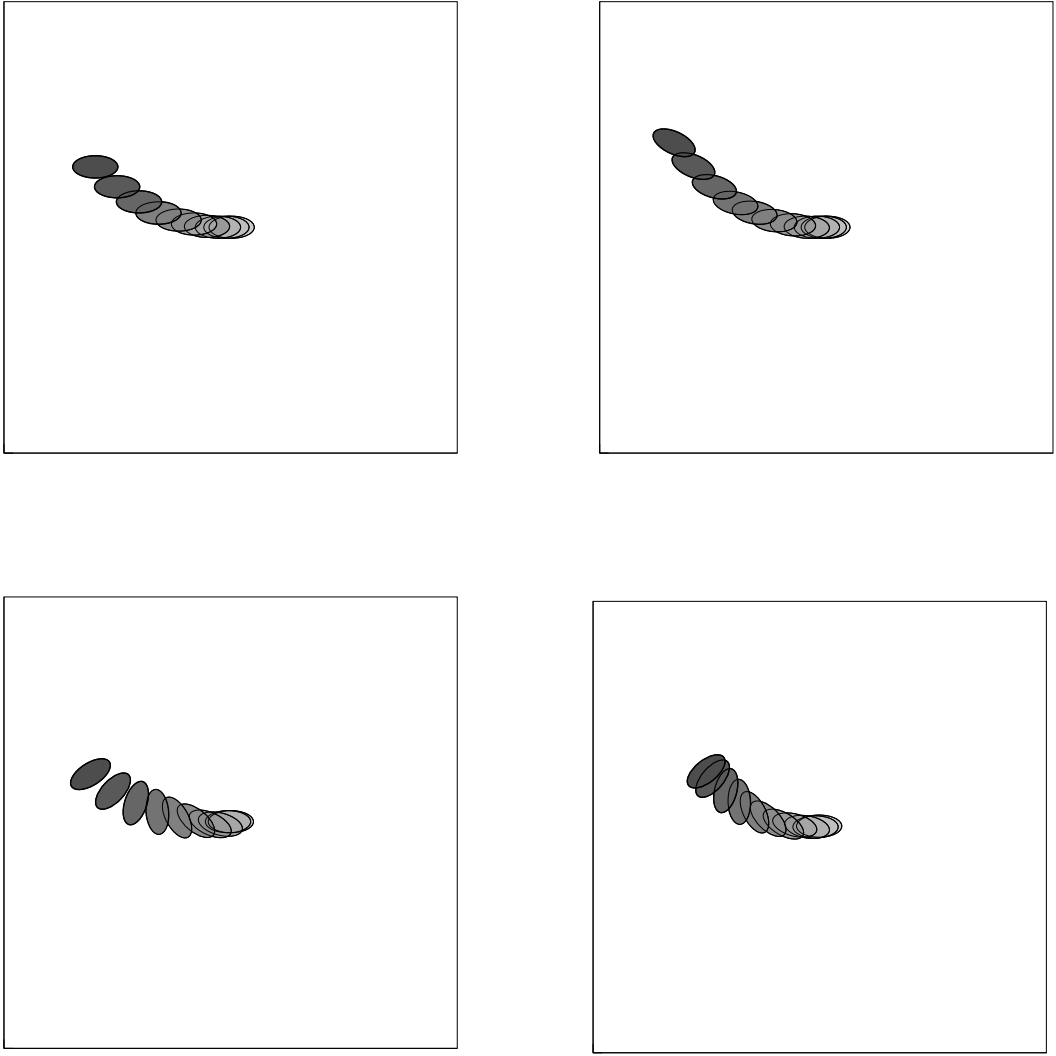}}
\end{center}
\caption{The trajectories of various included rigid bodies released
from rest at the centre of the driven cavity flow described. {\sc Top
Left:} $Re = 0.025$, ${\bar m} = 251.3$, ${\bar J}_{33} = 314.2$ and
$t = 3.6$ secs. {\sc Top Right:} $Re = 0.025$, ${\bar m} = 251.3$,
${\bar J}_{33} = 1.0$ and $t = 4.0$ secs. {\sc Bottom Left:} $Re =
0.025$, ${\bar m} = 251.3$, ${\bar J}_{33} = 0.1$ and $t = 3.6$ secs.
{\sc Bottom Right:} $Re = 1$, ${\bar m} = 1$, moment of inertia
(scaled) $= 0.1 $ and $t = 2.0$ secs.} \label{108}
\end{figure}

Various rigid bodies were introduced to the driven cavity flow
problem described in Subsection \ref{32}, in the absence of a body
force. The results in Fig. \ref{108} involve the ellipse
\[
\frac{x_1^2}{2^2} + \frac{x_2^2}{1^2} = 0.025^2,
\]
whose major axis is $0.1$. The quantities ${\bar m}$ and ${\bar
J}_{ii \mbox{\scriptsize (no sum)}}$ are a dimensionless mass and
$i$th principle moment of inertia respectively.

\subsection{Example 3: Die Swell Problems}

The axis--symmetric die swell (or fluid jet) problem is a free
surface problem well documented in the literature ({\sc Kruyt,
Cuvelier, Segal} and {\sc Van Der Zanden} \cite{k:1}, {\sc Omodei}
\cite{o:1} and {\sc Engelman} and {\sc Dupret} quoted in {\sc Kruyt
et al.}). The basic theme to this problem is the extrusion of a fluid
with initial parabolic flow profile from the end of a short nozzle. 
\begin{figure}[H]
\begin{center} \leavevmode
\mbox{\epsfbox{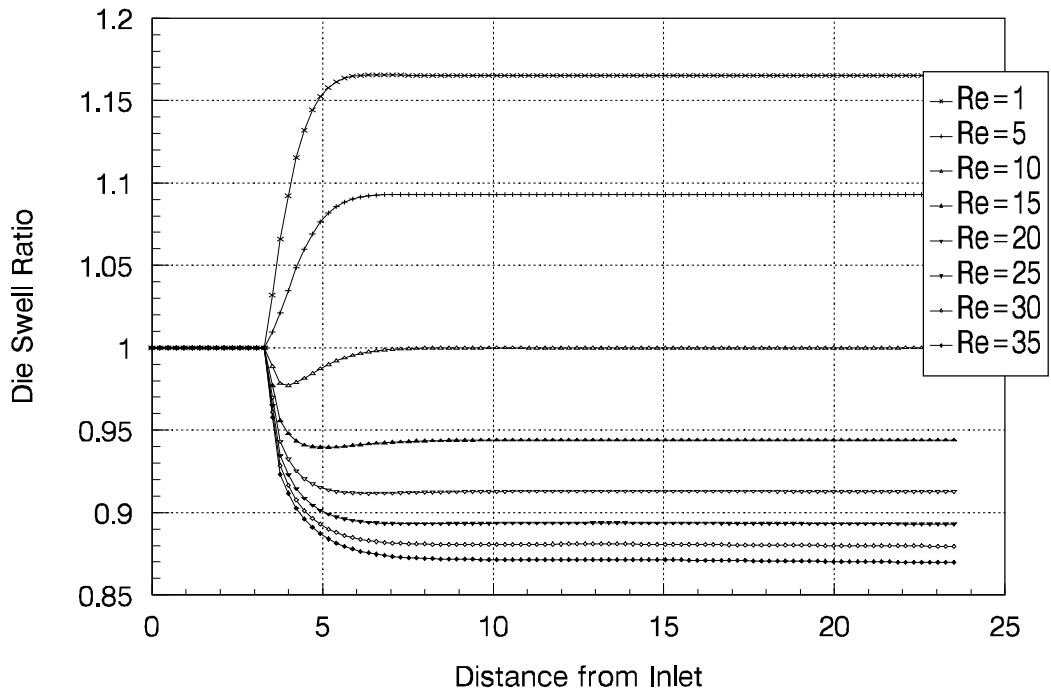}}
\end{center}
\caption{Die swell ratios predicted for various Reynolds numbers using
an inlet velocity profile of ${\bar v}_1 = \frac{Re}{1} \frac{3}{2}
(1-{{\bar x}_2^2})$ and the methods described. (Bars on the variables merely indicate that they are dimensionless.)} \label{131}
\end{figure}

\begin{figure}[H]
\begin{center} \leavevmode%%BoundingBox: 55 195 576 430
\mbox{\epsfbox{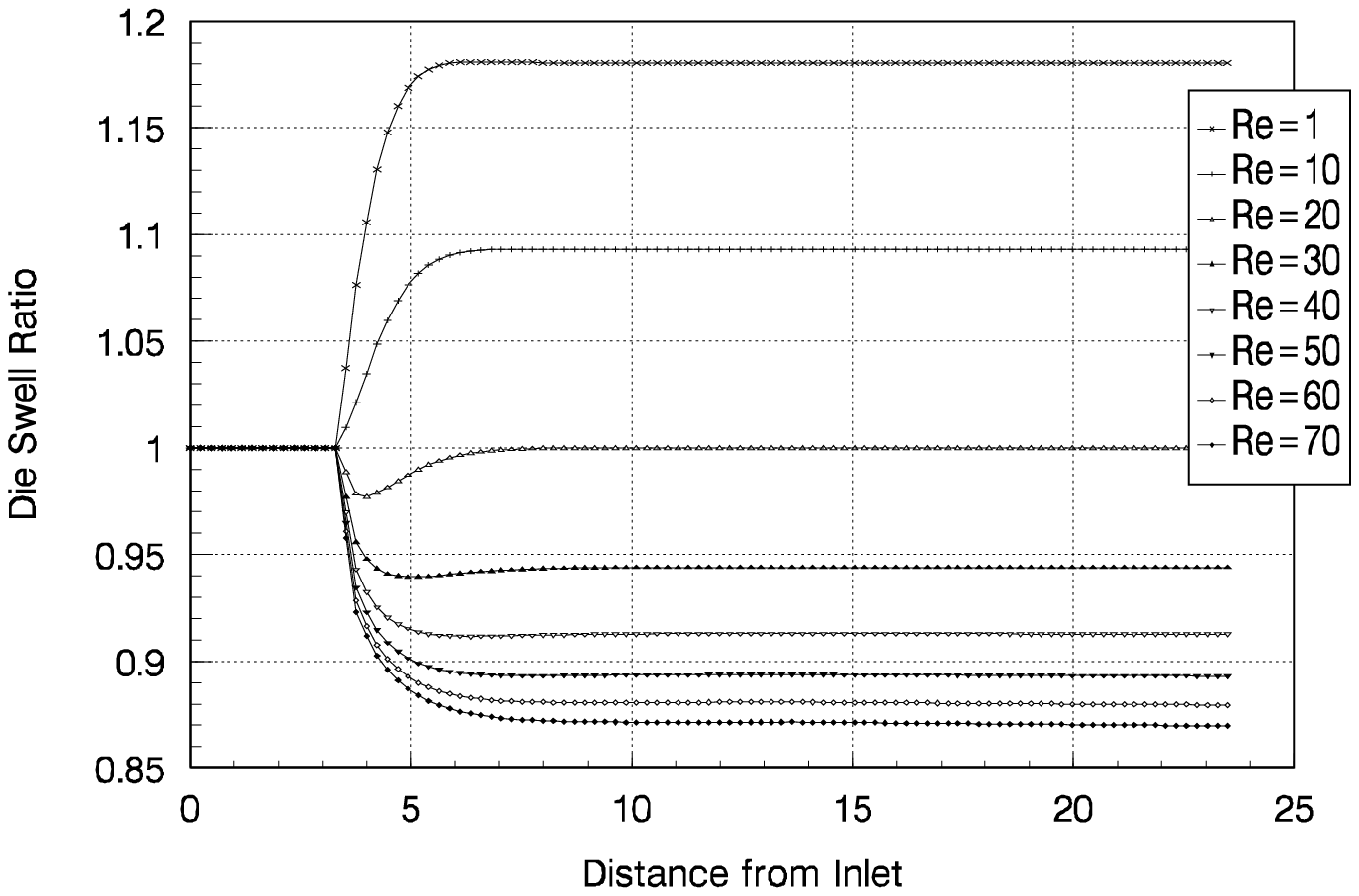}}
\end{center}
\caption{Die swell ratios predicted for various Reynolds numbers using
an inlet velocity profile of ${\bar v}_1 = \frac{Re}{10} \frac{3}{2}
(1-{{\bar x}_2^2})$ and the methods described.} \label{132}
\end{figure}

\begin{figure}[H]
\begin{center} \leavevmode
\mbox{\epsfbox{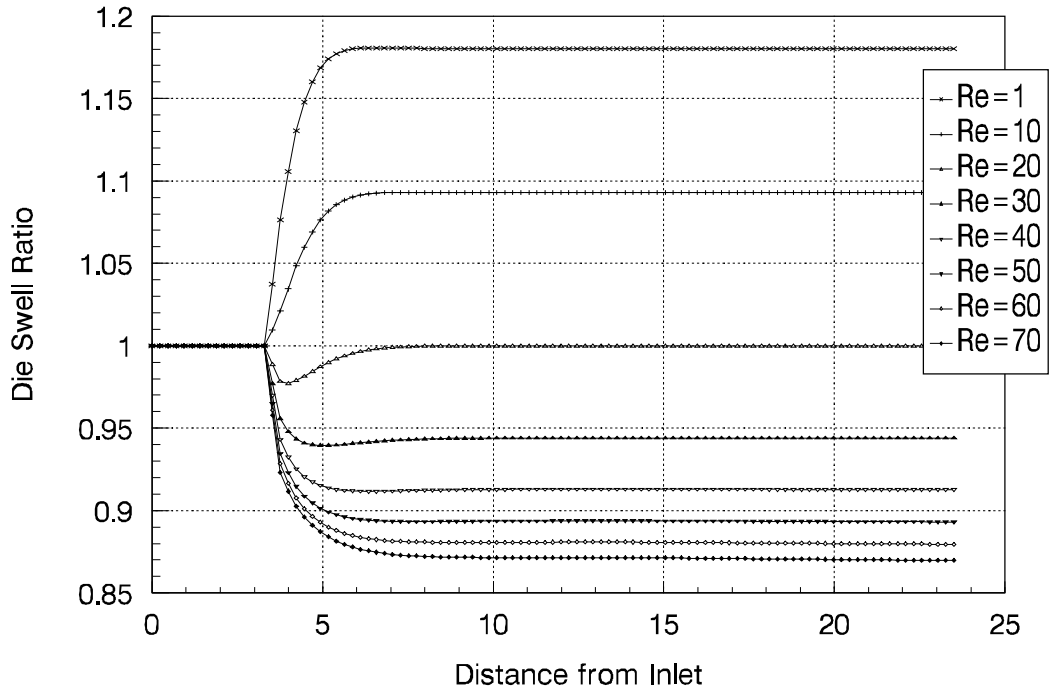}}
\end{center}
\caption{Die swell ratios predicted for various Reynolds numbers using
an inlet velocity profile of ${\bar v}_1 = \frac{Re}{20} \frac{3}{2}
(1-{{\bar x}_2^2})$ and the methods described.} \label{64}
\end{figure}

\subsection{Example 4: A Stokes Second Order Wave}

In this problem the velocity profile and surface elevation predicted
by Stokes second order wave theory (see {\sc Koutitas}
\cite{koutitas:1}) were used as boundary conditions for flow and free
surface subproblems respectively. 
\begin{figure}[H]
\begin{center} \leavevmode
\mbox{\epsfbox{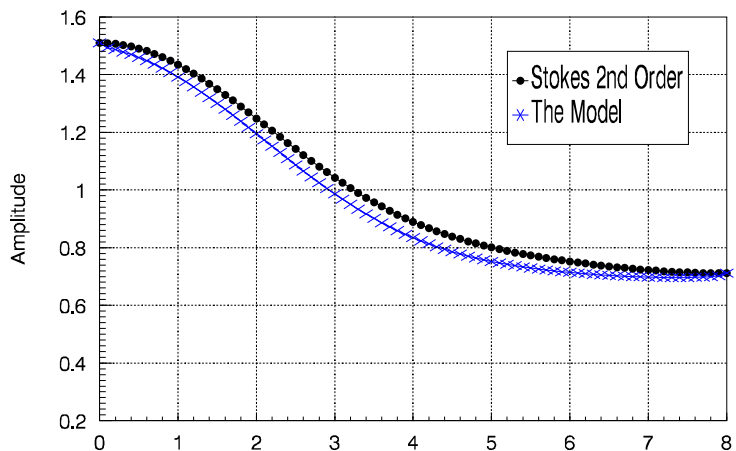}}
\end{center}
\caption{A Stokes second order Wave.}
\label{87}
\end{figure}
Problems with wave propagation were subsequently experienced as time
progressed. It should be noted, however, that the problem was not
attempted with the same seriousness as previous examples and the mesh
was poor (there being only three elements in the vertical extent of
the mesh).

\section{Conclusions}

The correct equations, which describe the motion of an
incompressible, Newtonian fluid and which are valid for a completely
general range of reference deformations, are equations (\ref{29}) and
(\ref{30}). For implementations requiring the equations to be
evaluated about a single instant within each time step only (eg.
finite differences), the deformation gradients may be assumed
identity i.e. the equations of {\sc Hughes, Liu} and {\sc Zimmerman}
\cite{h:1} (equations (\ref{307}) and (\ref{308})) will suffice.

In this work it is shown (as was hoped) that nonlinear,
exponential--type dissipation in the absence of forcing and
long--term stability under conditions of time dependent loading are
properties automatically inherited by such deforming reference
descriptions. The single provisor is that the conventional boundary
descriptions are used (vanishing ${\tilde {\bn}} \cdot \left( {\tilde
{\bv}} - {\tilde {\bv}}^{\mbox{\scriptsize {\em ref}}} \right)$ type
at free boundaries, purely Eulerian at boundaries across which there
is an imposed velocity or where boundaries are of a fixed,
impermeable type). These properties are intrinsic to real flows and
the conventional, Eulerian Navier--Stokes equations. 

Relevant energy terms are, however, not readily recovered from the
time--discrete equations for deforming references in general. Only
for divergence free rates of reference deformation could it
consequently be guaranteed that energy would not be artificially
introduced to the algorithmic flow by way of the reference. A further
casualty of the discrete analysis is its failure to account for flows
driven by their boundaries in any obvious way i.e. boundaries of an
imposed velocity type do not enter explicitly into the bound. Scope
for the further development of this work therefore exists.

The divergence free assumption was made for reasons of expedience
alone and it is hoped that the findings of the time--discrete
analysis can be extrapolated to a more general class of mesh
deformations. If one were to be overly cautious on this basis one
would be faced with the additional challenge of enforcing mesh
deformations which are divergence free. Such a totally divergence
free description may, however, not be practical. Both the purely
Lagrangian and purely Eulerian fluid descriptions have divergence
free rates of distortion.

What is clear is that there are inherent problems with using certain
classes of time--stepping schemes and the use of finite difference
schemes more implicit than central difference is consequently
advocated. The limitations of the time--discrete analysis do not
detract from this finding in any way. Such differences exhibit the
key energetic properties (nonlinear, exponential--type dissipation in
the absence of forcing and long--term stability under conditions of
time dependent loading) irrespective of the time increment employed.
A backward difference is the obvious choice. Calculations at time $t
+ \alpha \Delta t$ would require an intermediate mesh and associated
quantities for instances in which $\alpha \ne 1$ (since $\alpha >
\frac{1}{2}$).

The author recommends a strategy in which a predominantly Eulerian
description is used, where possible, for the bulk of the problem (from
an efficiency point of view) and the completely general reference
description for the remainder is appropriate. Purely Eulerian
descriptions have the advantage of a ``one off'' finite element
construction and involve none of the hazards of a badly distorted
reference.

With regard to numerical implementation and using a $Q_2$--$P_1$ element pair, it was found that pressures approximated as linear on
the master element still led to so--called ``locking'' or
``chequerboard'' modes. The pressures needed to be linear on the
actual elements themselves. This finding makes sense if one considers
that a linear function mapped from the master element using a $Q_2$
mapping will no longer be $P_1$ for non--rectangular elements (the
$Q_2$--$P_1$ element pair was shown to satisfy the L.B.B. condition
in the context of rectangular elements).

Lastly, the linearised terms $( 2 {\bv} \mid_t - {\bv} \mid_{t-\Delta
t} ) \cdot \nabla {\bv} \mid_{t+\Delta t} $ and $ {\bv}
\mid_{t+\Delta t} \cdot \nabla (2 {\bv} \mid_t - {\bv} \mid_{t-\Delta
t})$ are second order accurate approximations of the convective term
and a remarkably practical, simple and effective method to
automatically generate meshes about included rigid bodies was devised.

\section{Acknowledgements}

Daya Reddy is thanked for the loan of references and related advice.
Grzegorz Lubczonok and Ronald Becker are thanked for their respective
opinions on the inequality (Inequality \ref{94}). Others who assisted the
author in various ways (including the provision of resources and moral
support) are Kevin Colville, George Ellis and Rait Harnett. The final
arrangement and editing of this work was carried out under the
supervision of Philip Gresho.

\section{Appendix I}

\subsection{The Contribution ${\tilde K}_{{\tilde \Gamma} \mbox{\tiny imposed } \mbox{\tiny ${\tilde {\bv}}$}}$ at Imposed Velocity--Type Boundaries}

Additional terms which arise from the limits of the integral 
\[
{\rho} \int_{{\tilde x}_1^0}^{{\tilde x}_1'} \int_{{\tilde
x}_2^0}^{{\tilde x}_2'} \int_{{\tilde x}_3^0}^{{\tilde x}_3'} {\tilde
{\bv}} \cdot \frac{\partial {\tilde {\bv}}}{\partial t} {\tilde J} \ 
d{\tilde x}_3 d{\tilde x}_2 d{\tilde x}_1
\]
when changing the order of differentiation and integration at boundaries across which there is an imposed velocity are:
\begin{eqnarray} \label{43}
&& - \frac{1}{2}{\rho} \left[ \frac{D {\tilde x}_1'}{Dt}
\left.\left(\int_{{\tilde x}_2^0}^{{\tilde x}_2'} \int_{{\tilde
x}_3^0}^{{\tilde x}_3'} {\tilde {\bv}} \cdot {\tilde {\bv}} {\tilde
J} d{\tilde x}_3 d{\tilde x}_2\right)\right|_{{\tilde x}_1'} -
\frac{D {\tilde x}_1^0}{Dt} \left.\left(\int_{{\tilde
x}_2^0}^{{\tilde x}_2'} \int_{{\tilde x}_3^0}^{{\tilde x}_3'} {\tilde
{\bv}} \cdot {\tilde {\bv}} {\tilde J} d{\tilde x}_3 d{\tilde
x}_2\right)\right|_{{\tilde x}_1^0} \right. \nonumber \\
&& + \int_{{\tilde x}_1^0}^{{\tilde x}_1'} \frac{D {\tilde x}_2'}{Dt}
\left.\left(\int_{{\tilde x}_3^0}^{{\tilde x}_3'} {\tilde {\bv}}
\cdot {\tilde {\bv}} {\tilde J} d{\tilde x}_3\right)\right|_{{\tilde
x}_2'} d{\tilde x}_1 - \int_{{\tilde x}_1^0}^{{\tilde x}_1'} \frac{D
{\tilde x}_2^0}{Dt} \left.\left(\int_{{\tilde x}_3^0}^{{\tilde x}_3'}
{\tilde {\bv}} \cdot {\tilde {\bv}} {\tilde J} d{\tilde
x}_3\right)\right|_{{\tilde x}_2^0} d{\tilde x}_1 \nonumber \\
&& \left. + \int_{{\tilde x}_1^0}^{{\tilde x}_1'} \int_{{\tilde
x}_2^0}^{{\tilde x}_2'} \frac{D {\tilde x}_3'}{Dt}
\left.\left({\tilde {\bv}} \cdot {\tilde {\bv}} {\tilde
J}\right)\right|_{{\tilde x}_3'} d{\tilde x}_2 d{\tilde x}_1 -
\int_{{\tilde x}_1^0}^{{\tilde x}_1'} \int_{{\tilde x}_2^0}^{{\tilde
x}_2'} \frac{D {\tilde x}_3^0}{Dt} \left.\left({\tilde {\bv}} \cdot
{\tilde {\bv}} {\tilde J}\right)\right|_{{\tilde x}_3^0} d{\tilde
x}_2 d{\tilde x}_1 \right] 
\end{eqnarray}
(using Leibnitz's rule for differentiation under the integral sign).

Note that the terms $\displaystyle \frac{D \ldots}{Dt}$ can be given
a 
\[
\frac{\partial \ldots}{\partial t} + {\tilde \nabla} \ldots {\tilde {\bF}}^{-1} (\tilde{\bv} - \tilde{\bv}^{ref}) 
\] 
interpretation in terms of the relation (\ref{25}) established
at the beginning of Section \ref{2}. For descriptions which are
purely Eulerian at such fixed boundaries, ${\tilde{\bv}}^{ref}$
vanishes and ${\tilde {\bF}}$ is identity. Thus the extra terms, (\ref{43}) above, can be rewritten (with minus sign omitted) 
\begin{eqnarray*}
&& \frac{1}{2}{\rho} \left[ \nabla x'_{1} \cdot {\bv}
\left.\left(\int_{x_2^0}^{x_2'} \int_{x_3^0}^{x_3'} {\bv} \cdot {\bv}
\ dx_3 dx_2\right)\right|_{{\tilde x}_1'} - \nabla x^0_{1} \cdot
{\bv} \left.\left(\int_{x_2^0}^{x_2'} \int_{x_3^0}^{x_3'} {\bv} \cdot
{\bv} \ dx_3 dx_2\right)\right|_{{\tilde x}_1^0} \right. \\
&& + \int_{x_1^0}^{x_1'} \nabla x'_{2} \cdot {\bv}
\left.\left(\int_{x_3^0}^{x_3'} {\bv} \cdot {\bv} \
dx_3\right)\right|_{{\tilde x}_2'} dx_1 - \int_{x_1^0}^{x_1'} \nabla
x^0_{2} \cdot {\bv} \left.\left(\int_{x_3^0}^{x_3'} {\bv} \cdot {\bv}
\ dx_3\right)\right|_{{\tilde x}_2^0} dx_1 \\
&& \left. + \int_{x_1^0}^{x_1'} \int_{x_2^0}^{x_2'} \nabla x'_{3}
\cdot {\bv} \left.\left({\bv} \cdot {\bv}\right)\right|_{{\tilde
x}_3'} \ dx_2 dx_1 - \int_{x_1^0}^{x_1'} \int_{x_2^0}^{x_2'} \nabla
x^0_{3} \cdot {\bv} \left.\left({\bv} \cdot
{\bv}\right)\right|_{{\tilde x}_3^0} \ dx_2 dx_1 \right] 
\end{eqnarray*}
i.e. a total rate of energy transport across the boundaries, similar, but not identical to $\int_{\Gamma} ({\bn} \cdot {\bv}) ({\bv} \cdot {\bv}) d \Gamma$. These terms, dubbed ${\tilde K}_{{\tilde \Gamma} \mbox{\tiny imposed } \mbox{\tiny ${\tilde {\bv}}$}}$ in the work, must be added to the right hand side of equation (\ref{59}) when circumstances require.

This approach may seem comparatively crude in the light of rather
elegant work done by {\sc Temam} \cite{Temam:2} for flows driven by
their boundaries, however, the intended purpose differs slightly.
What is here being sought is a bound formulated in terms of known,
physically comprehensible quantities at the boundary which are
independant of the solution. 

\subsection{The Ramifications of Assumption \ref{assumption}} \label{45}

If one does not make Assumption \ref{assumption}, contributions from the 
\[
\frac{\rho}{\Delta t} \left< {\tilde {\bv}}_{n+\alpha}, ({\tilde {\bv}}_{n+1} - {\tilde {\bv}}_{n}) {\tilde J}_{n+\alpha} \right>
\]
term in equation (\ref{70}) and 
\[
\frac{1}{2} \rho \left< {\tilde {\bv}}_{n+\alpha}, {\tilde
{\bv}}_{n+\alpha} \frac{({\tilde J}_{n+1} - {\tilde J}_{n})}{\Delta
t}\right>
\]
in Lemma \ref{355} amount to
\begin{eqnarray*}
&& \frac{\rho}{\Delta t} \left< {\tilde {\bv}}_{n+\alpha}, ({\tilde {\bv}}_{n+1} - {\tilde {\bv}}_{n}) {\tilde J}_{n+\alpha} \right> + \frac{\rho}{\Delta t} \frac{1}{2} \left< {\tilde {\bv}}_{n+\alpha}, {\tilde {\bv}}_{n+\alpha} ({\tilde J}_{n+1} - {\tilde J}_{n})\right> \\
&& \hspace{10mm}= \frac{\rho}{\Delta t} \frac{1}{(\alpha -
\frac{1}{2})}\left[ \left< {\tilde {\bv}}_{n+\alpha}, ({\tilde
{\bv}}_{n+\alpha} - {\tilde {\bv}}_{n + \frac{1}{2}}) {\tilde
J}_{n+\alpha} \right> + \frac{1}{2}\left< {\tilde {\bv}}_{n+\alpha},
{\tilde {\bv}}_{n+\alpha} ({\tilde J}_{n+\alpha} - {\tilde
J}_{n+\frac{1}{2}})\right> \right] \\
&& \hspace{10mm}= \frac{\rho}{\Delta t} \frac{1}{(\alpha -
\frac{1}{2})}\left[ \frac{3}{2}\left< {\tilde {\bv}}_{n+\alpha},
{\tilde {\bv}}_{n+\alpha} {\tilde J}_{n+\alpha} \right> - \left< {\tilde {\bv}}_{n+\alpha},
{\tilde {\bv}}_{n+\frac{1}{2}} {\tilde J}_{n+\alpha} \right> -
\frac{1}{2}\left< {\tilde {\bv}}_{n+\alpha}, {\tilde
{\bv}}_{n+\alpha} {\tilde J}_{n+\frac{1}{2}}\right> \right] \\
&& \\
&& \hspace{10mm}= \frac{\rho}{\Delta t} \frac{1}{(\alpha -
\frac{1}{2})}\left[ \frac{3}{2}\left< \alpha{\tilde {\bv}}_{n+1} +
\left(1-\alpha\right){\tilde {\bv}}_{n}, \alpha{\tilde {\bv}}_{n+1} +
\left(1-\alpha\right){\tilde {\bv}}_{n} \alpha{\tilde J}_{n+1} +
\left(1-\alpha\right){\tilde J}_{n} \right> \right. \\
&& \hspace{45mm} - \left< \alpha{\tilde {\bv}}_{n+1} +
\left(1-\alpha\right){\tilde {\bv}}_{n}, \frac{1}{2}\left({\tilde
{\bv}}_{n+1} + {\tilde {\bv}}_{n}\right) \alpha{\tilde J}_{n+1} +
\left(1-\alpha\right){\tilde J}_{n} \right> \\
&& \hspace{67mm} - \left. \frac{1}{2}\left< \alpha{\tilde {\bv}}_{n+1} +
\left(1-\alpha\right){\tilde {\bv}}_{n}, {\tilde {\bv}}_{n+\alpha}
\frac{1}{2}\left({\tilde J}_{n+1} + {\tilde J}_{n} \right)\right>
\right] \\
&& \\
&& \hspace{10mm}= 3\alpha^2\frac{1}{\Delta t}\left({\tilde K}({\tilde {\bv}}_{n+1}) -
{\tilde K}({\tilde {\bv}}_{n})\right) + \frac{1}{\Delta t}6\left(\alpha -
\frac{1}{2}\right){\tilde K}({\tilde {\bv}}_{n}) \\
&& \hspace{15mm} + \frac{\rho}{\Delta t} \left[ \frac{\alpha}{2}(2 -
3\alpha)\left<{\tilde {\bv}}_{n + 1}, {\tilde {\bv}}_{n +
1}J_n\right> \right. + \alpha(2 - 3\alpha)\left<{\tilde {\bv}}_{n},
{\tilde {\bv}}_{n + 1}J_{n+1}\right> \\
&& \hspace{15mm} - \frac{1}{4}(1 - \alpha)(3\alpha - 1)\left<{\tilde
{\bv}}_{n}, {\tilde {\bv}}_{n + 1}J_{n}\right> \left. + \frac{(1 -
\alpha)^2(6\alpha - 1)}{4(\alpha - \frac{1}{2})} \left<{\tilde
{\bv}}_{n}, {\tilde {\bv}}_{n}J_{n+1}\right> \right] \\
\end{eqnarray*}
(by repeated substitution of (\ref{52}) and (\ref{48})). Thus the ramifications of Assumption \ref{assumption} are that the total  
\begin{eqnarray*} 
&& \hspace{-5mm} \frac{1}{\Delta t}6\left(\alpha -
\frac{1}{2}\right){\tilde K}({\tilde {\bv}}_{n}) + \frac{\rho}{\Delta
t} \left[ \frac{\alpha}{2}(2 - 3\alpha)\left<{\tilde {\bv}}_{n + 1},
{\tilde {\bv}}_{n + 1}J_n\right> \right. + \alpha(2 -
3\alpha)\left<{\tilde {\bv}}_{n}, {\tilde {\bv}}_{n +
1}J_{n+1}\right> \\
&& \hspace{30mm} - \frac{1}{4}(1 - \alpha)(3\alpha - 1)\left<{\tilde
{\bv}}_{n}, {\tilde {\bv}}_{n + 1}J_{n}\right> \left. + \frac{(1 -
\alpha)^2(6\alpha - 1)}{4(\alpha - \frac{1}{2})} \left<{\tilde
{\bv}}_{n}, {\tilde {\bv}}_{n}J_{n+1}\right> \right] \\
\end{eqnarray*}
is positive, furthermore it is sufficiently positive to offset any subsequent short--coming which arises when Assumption \ref{assumption} is exploited in the proof of Lemma \ref{57} i.e. in the event of 
\[
4\nu C \alpha(1-\alpha)\left(\left|\left| {\tilde {\bv}}_{n+1} {\tilde J}_{n+1} \right|\right|^2 - \left|\left| {\tilde {\bv}}_{n+1} {\tilde J}_{n+\alpha} \right|\right|^2\right)
\]
and
\[
4\nu C \alpha(1-\alpha)\left(\left|\left| {\tilde {\bv}}_{n} {\tilde J}_{n} \right|\right|^2 - \left|\left| {\tilde {\bv}}_{n} {\tilde J}_{n+\alpha} \right|\right|^2\right)
\]
not being positive. 

\section{Appendix II (Proofs)}

\subsubsection*{Proof of Relation \ref{25}}

The above relation (taken from {\sc Hughes, Liu} and {\sc Zimmerman}
\cite{h:1}) is obtained by recalling that the material derivative
(total derivative) is the derivative with respect to time in the
material configuration. Thus
\begin{eqnarray} \label{26}
\frac{D\tilde{v}_i}{Dt} & = & \frac{\partial}{\partial t} \{ {\tilde
v}_i ( {\tilde {{\blambda}}} ({\bx} _0,t) ,t)  \} \nonumber \\
& = & \frac{\partial \tilde{v}_i}{\partial t} + \frac{\partial
\tilde{v}_i}{\partial \tilde{x}_j}\frac{\partial
\tilde{\lambda}_j}{\partial t} \ .
\end{eqnarray}
A more practical expression is needed for $\displaystyle \frac
{\partial\tilde{\lambda_j}} {\partial t}$ (the velocity as perceived in the distorting reference). This can be obtained by considering
\[
\lambda_k({\bx} _0,t) = \lambda^{\ast}_k(\tilde {{\blambda}} ({\bx}
_0,t),t) \hspace{10mm} \mbox{(see Fig. \ref{91})}
\] 
so that
\[ 
{\left.\frac{ \partial \lambda_k }{\partial t} \right|} _{ {\bx} _0
\ fixed} = {\left.\frac{ \partial \lambda^{\ast}_k }{\partial t}
\right|}_{ {\tilde{\bx} } \ fixed} + \frac{ \partial \lambda^{\ast}_k
}{\partial \tilde{x}_j}\frac{\partial \tilde{\lambda}_j}{\partial t}
\]
or
\[ 
\frac{\partial \tilde{\lambda}_j}{\partial t} = \frac{\partial
{\tilde x}_j}{\partial x_k} \left( {\left. \frac{ \partial \lambda_k
}{\partial t} \right|}_{ {\bx} _0 \ fixed} - {\left. \frac{ \partial
\lambda^{\ast}_k }{\partial t} \right|}_{ {\tilde {\bx} } \ fixed}
\right).
\] 
Substituting this expression into equation (\ref{26}), the desired,
suitably practicable result is obtained.

\subsubsection*{Proof of Inequality \ref{94}} 

Consider the change to spherical coordinates
\[
{\breve v}_i(r, \theta, \phi) = { v}_i(r \mathop{\rm sin}\theta
\mathop{\rm cos}\phi - { x}_1^{\mbox{\scriptsize origin}}, r
\mathop{\rm sin}\theta \mathop{\rm sin}\phi - {
x}_2^{\mbox{\scriptsize origin}}, r \mathop{\rm cos}\theta -
{ x}_3^{\mbox{\scriptsize origin}})
\]
centred on ${{\bx}}^{\mbox{\scriptsize origin}}$. Suppose the radial
limits of the domain and neighbourhood are denoted $R_b (\theta, \phi)$
and $R_a (\theta, \phi)$ respectively. By the fundamental theorem of
integral calculus
\begin{eqnarray*}
\left( {\breve v}_i(r, \theta, \phi) - {\breve v}_i\mid_{R_a (\theta,
\phi)} \right)^2 &=& \left( \int_{R_a (\theta, \phi)}^r \frac{\partial
{\breve v}_i}{\partial r} (\xi, \theta, \phi) d \xi \right)^2 \\ 
&& \\
&=& \left( \int_{R_a (\theta, \phi)}^{r} \frac{1}{\xi} \xi
\frac{\partial {\breve v}_i}{\partial r} (\xi, \theta, \phi) d \xi
\right)^2 \\
&& \\
&\le& \int_{R_a (\theta, \phi)}^r \frac{1}{\xi^2} d \xi \int_{R_a
(\theta, \phi)}^r \left( \frac{\partial {\breve v}_i}{\partial r}(\xi,
\theta, \phi) \right)^2 \xi^2 d \xi \\
&& \hspace{10mm} \mbox{\it (by Schwarz inequality)} \\
%&& \\
&\le& \int_{R_{\mbox{\it \scriptsize min}}}^{R_{\mbox{\it \scriptsize
max}}} \frac{1}{\xi^2} d \xi \int_{R_a (\theta, \phi)}^{R_b (\theta,
\phi)} \left( \frac{\partial {\breve v}_i}{\partial {r}}(\xi,
\theta, \phi) \right)^2 \xi^2 d \xi \hspace{10mm} \mbox{\it (for } r
\in {\breve \Omega} \mbox{\it)} \\
&& \\
&=& \frac{(R_{\mbox{\it \scriptsize max}} - R_{\mbox{\it \scriptsize
min}})}  {R_{\mbox{\it \scriptsize max}} R_{\mbox{\it \scriptsize
min}}} \int_{R_a (\theta, \phi)}^{R_b (\theta, \phi)} \left(
\frac{\partial {\breve v}_i}{\partial r}(\xi, \theta, \phi)
\right)^2 \xi^2 d \xi \\
&& \\ 
&=& \frac{(R_{\mbox{\it \scriptsize max}} - R_{\mbox{\it \scriptsize
min}})}  {R_{\mbox{\it \scriptsize max}} R_{\mbox{\it \scriptsize
min}}} {\breve V}_i (\theta, \phi)
\end{eqnarray*}
\[
\mbox{where} \hspace{10mm} {\breve V}_i (\theta, \phi) = \int_{R_a
(\theta, \phi)}^{R_b (\theta, \phi)} \left( \frac{\partial {\breve
v}_i}{\partial r}(\xi, \theta, \phi) \right)^2 \xi^2 d \xi.
\]
Integrating this result over that part of ${\breve \Omega}$ outside the
neighbourhood (angular extent being $\Theta_a(\phi) \le \theta \le
\Theta_b(\phi)$ and $\Phi_a \le \phi \le \Phi_b$)
\begin{eqnarray*}
&& \hspace{-7mm} \int_{\Phi_a}^{\Phi_b}
\int_{\Theta_a(\phi)}^{\Theta_b(\phi)} \int_{R_a (\theta, \phi)}^{R_b
(\theta, \phi)} \left( {\breve v}_i(r, \theta, \phi) - {\breve
v}_i\mid_{R_a (\theta, \phi)} \right)^2 r^2 \mathop{\rm sin}\theta d r
d \theta d \phi \\
&& \hspace{13mm} \le \frac{(R_{\mbox{\it \scriptsize max}} -
R_{\mbox{\it \scriptsize min}})}  {R_{\mbox{\it \scriptsize max}}
R_{\mbox{\it \scriptsize min}}} \int_{\Phi_a}^{\Phi_b}
\int_{\Theta_a(\phi)}^{\Theta_b(\phi)} \int_{R_a (\theta, \phi)}^{R_b
(\theta, \phi)} {\breve V}_i (\theta, \phi) r^2 \mathop{\rm sin}\theta
d r d \theta d \phi \\
&& \\
&& \hspace{13mm} \le \frac{(R_{\mbox{\it \scriptsize max}} -
R_{\mbox{\it \scriptsize min}})}  {R_{\mbox{\it \scriptsize max}}
R_{\mbox{\it \scriptsize min}}} \int_{\Phi_a}^{\Phi_b}
\int_{\Theta_a(\phi)}^{\Theta_b(\phi)} {\breve V}_i (\theta, \phi)
\left( \int_{R_{\mbox{\it \scriptsize min}}}^{R_{\mbox{\it \scriptsize
max}}} r^2 d r \right) \mathop{\rm sin}\theta d \theta d \phi \frac{}{}
\\
&& \\
&& \hspace{13mm} \le \frac{(R_{\mbox{\it \scriptsize max}} -
R_{\mbox{\it \scriptsize min}}) (R_{\mbox{\it \scriptsize max}}^3 -
R_{\mbox{\it \scriptsize min}}^3)} {3 R_{\mbox{\it \scriptsize max}}
R_{\mbox{\it \scriptsize min}}} \int_{\Phi_a}^{\Phi_b}
\int_{\Theta_a(\phi)}^{\Theta_b(\phi)} \int_{R_a (\theta, \phi)}^{R_b
(\theta, \phi)}  \left( \frac{\partial {\breve v}_i}{\partial r}
\right)^2 {r}^2 \mathop{\rm sin}\theta d r d \theta d \phi \\
&& \\
&& \hspace{13mm} \le \frac{(R_{\mbox{\it \scriptsize max}} -
R_{\mbox{\it \scriptsize min}}) (R_{\mbox{\it \scriptsize max}}^3 -
R_{\mbox{\it \scriptsize min}}^3)} {3 R_{\mbox{\it \scriptsize max}}
R_{\mbox{\it \scriptsize min}}} \int_{\Phi_a}^{\Phi_b}
\int_{\Theta_a(\phi)}^{\Theta_b(\phi)} \int_{R_a (\theta, \phi)}^{R_b
(\theta, \phi)} \left[ \left( \frac{\partial {\breve v}_i}{\partial r}
\right)^2 + \frac{1}{r^2} \left( \frac{\partial {\breve v}_i}{\partial
\theta} \right)^2 \right. \\ && \hspace{83mm} \left. + \frac{1}{r^2
\sin^2 \theta} \left( \frac{\partial {\breve v}_i}{\partial
\phi} \right)^2 \right] r^2 \mathop{\rm sin}\theta d r d \theta d \phi
\\
&& \\
&& \hspace{13mm} = \frac{(R_{\mbox{\it \scriptsize max}} - R_{\mbox{\it
\scriptsize min}}) (R_{\mbox{\it \scriptsize max}}^3 - R_{\mbox{\it
\scriptsize min}}^3)} {3 R_{\mbox{\it \scriptsize max}} R_{\mbox{\it
\scriptsize min}}} \int_{\Phi_a}^{\Phi_b}
\int_{\Theta_a(\phi)}^{\Theta_b(\phi)} \int_{R_a (\theta, \phi)}^{R_b
(\theta, \phi)} \left( \nabla {\breve v}_i \right) \cdot \left( \nabla
{\breve v}_i \right) r^2 \mathop{\rm sin}\theta dr d \theta d \phi.
\end{eqnarray*}

Changing back to the original rectangular coordinates and defining
${{\bv}}\mid_{\mbox{\it \scriptsize bndry}}$ to be a radially constant
function throughout $\Omega$ which takes the values of ${\breve {\bv}}
\mid_{R_a (\theta, \phi)}$ for $r = R_a(\theta, \phi)$,
\begin{eqnarray*}
\int_{\Omega_*} \left( { v}_i({ {\bx}}) - { v}_i\mid_{\mbox{\it
\scriptsize bndry}} \right)^2 d { \Omega} &\le& \frac{(R_{\mbox{\it
\scriptsize max}} - R_{\mbox{\it \scriptsize min}}) (R_{\mbox{\it
\scriptsize max}}^3 - R_{\mbox{\it \scriptsize min}}^3)} {3
R_{\mbox{\it \scriptsize max}} R_{\mbox{\it \scriptsize min}}}
\int_{\Omega_*} \left( \nabla { v}_i({ {\bx}}) \right) \cdot \left(
\nabla { v}_i({ {\bx}}) \right) d {\Omega}
\end{eqnarray*}
where $\Omega_*$ is $\Omega$ excluding the neighbourhood. Summing over
$i$,
\begin{eqnarray*}
\int_{\Omega_*} \left( { {\bv}} - {\bv}\mid_{\mbox{\it \scriptsize
bndry}} \right) \cdot \left( { {\bv}} - { {\bv}}\mid_{\mbox{\it
\scriptsize bndry}} \right) d { \Omega} &\le& \frac{(R_{\mbox{\it
\scriptsize max}} - R_{\mbox{\it \scriptsize min}}) (R_{\mbox{\it
\scriptsize max}}^3 - R_{\mbox{\it \scriptsize min}}^3)} {3
R_{\mbox{\it \scriptsize max}} R_{\mbox{\it \scriptsize min}}}
\int_{\Omega_*} \left( \nabla { {\bv}} \right) : \left( \nabla { {\bv}}
\right) d {\Omega}.
\end{eqnarray*}
Making use of either the Cauchy--Schwarz or triangle inequality, 
\begin{eqnarray*}
\left( \left|\left| {\bv} \right|\right|_{L^2({\Omega_*})} - \left|\left|
{\bv}\mid_{\mbox{\it \scriptsize bndry}} \right|\right|_{L^2({\Omega_*})}
\right)^2 &\le& \frac{(R_{\mbox{\it \scriptsize max}} - R_{\mbox{\it
\scriptsize min}}) (R_{\mbox{\it \scriptsize max}}^3 - R_{\mbox{\it
\scriptsize min}}^3)} {3 R_{\mbox{\it \scriptsize max}} R_{\mbox{\it
\scriptsize min}}} \left|\left| \nabla {\bv} \right|\right|_{L^2({\Omega_*})}^2,
\end{eqnarray*}
and remembering that $\sup \mid {\breve {\bv}} \mid_{R_a (\theta,
\phi)} \mid \le c$,
\begin{eqnarray*}
\left|\left| {\bv} \right|\right|_{L^2({\Omega_*})} &\le& \left[
\frac{(R_{\mbox{\it \scriptsize max}} - R_{\mbox{\it \scriptsize min}})
(R_{\mbox{\it \scriptsize max}}^3 - R_{\mbox{\it \scriptsize min}}^3)}
{3 R_{\mbox{\it \scriptsize max}} R_{\mbox{\it \scriptsize min}}}
\right]^{\frac{1}{2}} \left|\left| { \nabla}{ {\bv}}
\right|\right|_{L^2({\Omega_*})} + \left|\left| c \right|\right|_{L^2({\Omega_*})}.
\end{eqnarray*}
Consider the terms $\left|\left| {\bv} \right|\right|_{L^2}$ and
$\left|\left| c \right|\right|_{L^2}$. Comparing these terms under
circumstances of $\sup \mid {\bv} \mid \le c$ leads to the conclusion
that the inequality holds over the neighbourhood and that the
inequality is therefore unaffected when the domain of integration is
extended to include the neighbourhood. Of course, the radial extension
of ${\bv}\mid_{\mbox{\it \scriptsize bndry}}$ can be used in place of
$c$ in instances where inclusion of the neighbourhood is not required.

\subsubsection*{Proof of Lemma \ref{58}} 

If, in particular, ${ {\bv}}\mid_{\mbox{\it \scriptsize
bndry}} = 0$ in Inequality \ref{94},
\begin{eqnarray*}
\left|\left| { {\bv}} \right|\right|_{L^2({ \Omega})} &\le&
\frac{\left|\left| { \nabla} { {\bv}} \right|\right|_{L^2({
\Omega})}}{\sqrt {C}} \\
C\frac{1}{2}\left|\left| { {\bv}} \right|\right|_{L^2({ \Omega})}^2
&\le& \left|\left| { {\bD}}({ {\bv}}) \right|\right|_{L^2({ \Omega})}^2
\end{eqnarray*}
(The relationship between ${\bD}$ and ${\nabla} {\bv}$ arises in the
context of the original equations involving $\mathop{\rm
div}{\bsigma}$. It is because \begin{eqnarray*}
\begin{array}{ccll} D_{ij,j} &=& \frac{1}{2} \left( v_{i,jj} + v_{j,ij}
\right) & \\
&=& \frac{1}{2} \left( v_{i,jj} + v_{j,ji} \right)
\hspace{10mm} & \mbox{\em (changing the order of differentiation)} \\
&=& \frac{1}{2} v_{i,jj} & \mbox{\em (${\mathop {\rm div}}{\bv} = 0$ by incompressibility}), \end{array} \end{eqnarray*}
assuming, of course, that ${\bv}$ is continuous and differentiable to
first order.) Rewriting in terms of ${\tilde \Omega}$
\begin{eqnarray*}
\frac{C}{\rho} {\tilde K}({\tilde {\bv}}) &\le& \left|\left| {\tilde
{\bD}}({\tilde {\bv}}) {\tilde J}^{\frac{1}{2}}
\right|\right|_{L^2({\tilde \Omega})}^2.
\end{eqnarray*}

\subsubsection*{Proof of Lemma \ref{355}} 

Consider ${\tilde{\bu}} \cdot ({\tilde \nabla}
{\tilde{\bv}}) {\tilde {\bF}}^{-1} {\tilde {\bw}} {\tilde J}$:
\begin{eqnarray*}
{\tilde u}_i {\tilde v}_{i,j} {\tilde F}^{-1}_{jk}{\tilde w}_k {\tilde
J} &=& - {\tilde u}_{i,j} {\tilde v}_i {\tilde F}^{-1}_{jk} {\tilde
w}_k {\tilde J} - {\tilde u}_i {\tilde v}_i ({\tilde F}^{-1}_{jk}
{\tilde w}_k {\tilde J})_{,j} + ({\tilde u}_i {\tilde v}_i {\tilde
F}^{-1}_{jk} {\tilde w}_k {\tilde J})_{,j}
\end{eqnarray*}
by the product rule. In the terms arising from $({\tilde F}^{-1}_{jk}
{\tilde w}_k {\tilde J})_{,j}$, both ${\tilde F}^{-1}_{jk,j}$ and ${\tilde J}_{,j} {\tilde F}^{-1}_{jk}$ vanish under the condtions specified (in section \ref{1001}) for equations (\ref{307}) and (\ref{308}) to be a completely general reference description. Thus
\begin{eqnarray*}
{\tilde u}_i {\tilde v}_{i,j} {\tilde F}^{-1}_{jk} {\tilde w}_k {\tilde
J} &=& - {\tilde u}_{i,j} {\tilde v}_i {\tilde F}^{-1} _{jk} {\tilde
w}_k {\tilde J} - {\tilde u}_i {\tilde v}_i {\tilde F}^{-1}_{jk}
{\tilde w}_{k,j} {\tilde J} + ({\tilde u}_i {\tilde v}_i {\tilde
F}^{-1} _{jk} {\tilde w}_k {\tilde J})_{,j}.
\end{eqnarray*} 
Integrating over the domain ${\tilde {\Omega}}$ and applying the
divergence theorem,
\begin{eqnarray} \label{309}
\left<{\tilde{\bu}}, ({\tilde \nabla} {\tilde {\bv}}) {\tilde
{\bF}}^{-1} {\tilde {\bw}} {\tilde J}  \right>_{L^2(\tilde
\Omega)} &=& - \left< {\tilde {\bv}}, ({\tilde \nabla}{\tilde{\bu}})
{\tilde {\bF}}^{-1} {\tilde {\bw}} {\tilde J} 
\right>_{L^2(\tilde \Omega)} \nonumber \\
&& - \left< {\tilde{\bu}} \left( {\tilde \nabla}{\tilde {\bw}}:{\tilde
{\bF}}^{-t} \right), {\tilde {\bv}} {\tilde J} \right>_{L^2(\tilde
\Omega)} \nonumber \\
&& + \left< {\tilde{\bu}}, {\tilde {\bv}} \left( {\tilde {\bF}}^{-t}
{\tilde {\bN}} \cdot {\tilde {\bw}} \right) {\tilde J} 
\right>_{L^2(\tilde \Gamma)}. 
\end{eqnarray}
Thus the term
\begin{eqnarray*} \label{310}
2 \left< {\tilde {\bv}}, ({\tilde \nabla} {\tilde {\bv}}) {\tilde
{\bF}}^{-1} \left( {\tilde {\bv}} - {\tilde {\bv}}^{\mbox{\scriptsize
{\em ref}}} \right) {\tilde J} \right>_{L^2({\tilde \Omega})} &=&
- \left< {\tilde{\bv}} \left( {\tilde \nabla}\left( {\tilde {\bv}} -
{\tilde {\bv}}^{\mbox{\scriptsize {\em ref}}} \right):{\tilde
{\bF}}^{-t} \right), {\tilde {\bv}} {\tilde J} \right>_{L^2(\tilde
\Omega)} \nonumber \\
&& + \left< {\tilde {\bv}}, {\tilde {\bv}} \left( {\tilde {\bF}}^{-t}
{\tilde {\bN}} \cdot ({\tilde {\bv}} - {\tilde
{\bv}}^{\mbox{\scriptsize {\em ref}}}) \right) {\tilde J}
\right>_{L^2(\tilde \Gamma)} \nonumber \\
&& \nonumber \\
&=& \left< {\tilde{\bv}} \left( {\tilde \nabla}{\tilde
{\bv}}^{\mbox{\scriptsize {\em ref}}}:{\tilde {\bF}}^{-t} \right),
{\tilde {\bv}} {\tilde J} \right>_{L^2(\tilde \Omega)} \nonumber \\
&& + \left< {\tilde {\bv}}, {\tilde {\bv}} \left( {\tilde {\bF}}^{-t}
{\tilde {\bN}} \cdot ({\tilde {\bv}} - {\tilde
{\bv}}^{\mbox{\scriptsize {\em ref}}}) \right) {\tilde J}
\right>_{L^2(\tilde \Gamma)} \nonumber \\
&& \hspace{10mm} \mbox{\em (by incompressibility)} \nonumber \\
&& \nonumber \\
&=& \left< {\tilde{\bv}} \left( {\tilde \nabla}{\tilde
{\bv}}^{\mbox{\scriptsize {\em ref}}}:{\tilde {\bF}}^{-t} \right),
{\tilde {\bv}} {\tilde J} \right>_{L^2(\tilde \Omega)} \nonumber \\
&& + \left< {\tilde {\bv}}, {\tilde {\bv}} \left( {\tilde {\bF}}^{-t}
{\tilde {\bN}} \cdot ({\tilde {\bv}} - {\tilde
{\bv}}^{\mbox{\scriptsize {\em ref}}}) \right) {\tilde J}
\right>_{L^2(\tilde \Gamma)} \nonumber \\
&& \nonumber \\
&=& \left< {\tilde {\bv}}, {\tilde {\bv}} \frac{\partial {\tilde
J}}{\partial t} \right>_{L^2({\tilde \Omega})} \nonumber \\
&& + \left< {\tilde {\bv}}, {\tilde {\bv}} \left( {\tilde {\bF}}^{-t}
{\tilde {\bN}} \cdot ({\tilde {\bv}} - {\tilde
{\bv}}^{\mbox{\scriptsize {\em ref}}}) \right) {\tilde J}
\right>_{L^2(\tilde \Gamma)}
\end{eqnarray*}
since $\displaystyle \frac{\partial {\tilde J}}{\partial t} = {\tilde
J} div {\bv}^{\scriptsize ref}$ (which is ${\tilde J}{\tilde
\nabla}{\tilde {\bv}}^{\mbox{\scriptsize {\em ref}}}:{\tilde
{\bF}}^{-t}$) in the same vein as ${\dot {\cal J}}_0 = {\cal J}_0
\mathop{\rm div}{\bv}$ (the kinematic result used in Section \ref{130}).

Equation (\ref{309}) is vital to the deforming reference analysis in
particular. It forms the basis to this lemma and another (Lemma \ref{354}) concerned with the time discrete analysis.

\subsubsection*{Proof of Lemma \ref{41}} 
 
By referring to Fig. \ref{91} one observes that
\[
\lambda_j({\bx} _0,t) = \lambda^{\ast}_j(\tilde {{\blambda}} ({\bx}
_0,t),t)
\] 
and consequently that
\[ 
{\left.\frac{ \partial \lambda_j }{\partial t} \right|} _{ {\bx} _0
\ fixed} = {\left.\frac{ \partial \lambda^{\ast}_j }{\partial t}
\right|}_{ {\tilde{\bx} } \ fixed} + \frac{ \partial \lambda^{\ast}_j
}{\partial \tilde{x}_i}\frac{\partial \tilde{\lambda}_i}{\partial t}.
\]
That is
\[ 
{\tilde {\bF}}^{-1} ({\tilde {\bv}} - {\tilde
{\bv}}^{\mbox{\scriptsize {\em ref}}}) = \frac{\partial
\tilde{\blambda}}{\partial t}.
\] 
In other words ${\tilde {\bF}}^{-1} ({\tilde {\bv}} - {\tilde
{\bv}}^{\mbox{\scriptsize {\em ref}}})$ is the velocity perceived in
the deforming reference. This perceived velocity is tangent to the
free surface since the description was stipulated to be one in which
${\tilde {\bn}} \cdot \left( {\tilde {\bv}} - {\tilde
{\bv}}^{\mbox{\scriptsize {\em ref}}} \right)$ vanishes at free
surfaces. Remembering that ${\tilde {\bN}}$ is a surface normal as
defined in terms of this self--same reference,
\[
{\tilde {\bN}} \cdot {\tilde {\bF}}^{-1} \left( {\tilde {\bv}} -
{\tilde {\bv}}^{\mbox{\scriptsize {\em ref}}} \right) = 0
\hspace{5mm} \Rightarrow \hspace{5mm} - \frac{1}{2} \rho \left<
{\tilde{\bv}}, {\tilde {\bv}} \left( {\tilde {\bF}}^{-t} {\tilde
{\bN}} \cdot \left( {\tilde {\bv}} - {\tilde
{\bv}}^{\mbox{\scriptsize {\em ref}}} \right) \right) {\tilde J} 
\right>_{L^2(\tilde \Gamma)} = 0
\]
at free boundaries. 

\subsubsection*{Proof of Lemma \ref{49}} 

In terms of the Cauchy--Schwarz inequality,
\begin{eqnarray*} 
\left< {\tilde {\bv}}, {\tilde {\bb}} {\tilde J} \right>_{L^2({\tilde
\Omega})} &\le& \left|\left| {\tilde {\bv}} {\tilde J}^{\frac{1}{2}}
\right|\right|_{L^2({\tilde \Omega})} \ \left|\left| {\tilde {\bb}}
{\tilde J}^{\frac{1}{2}} \right|\right|_{L^2(\tilde \Omega)} \\
&\le& \frac{\nu C}{2} \left|\left| {\tilde {\bv}} {\tilde
J}^{\frac{1}{2}} \right|\right|_{L^2(\tilde \Omega)}^2 + \frac{1}{2 \nu
C} \left|\left| {\tilde {\bb}} {\tilde J}^{\frac{1}{2}}
\right|\right|^2_{L^2(\tilde \Omega)} \hspace{5mm} \mbox{for }
\hspace{5mm} \nu C > 0
\end{eqnarray*}
by Young's inequality. Similarly,
\begin{eqnarray*}
\left< {\tilde {\bv}}, {\tilde {\bP}} {\tilde {\bN}}
\right>_{L^2(\tilde \Gamma)} &\le& \frac{\nu C}{2} \left|\left| {\tilde
{\bv}} \right|\right|_{L^2({\tilde \Gamma})}^2 + \frac{1}{2 \nu C}
\left|\left| {\tilde {\bP}} {\tilde {\bN}} \right|\right|^2_{L^2(\tilde
\Gamma)} \hspace{5mm} \mbox{for } \hspace{5mm} \nu C > 0.
\end{eqnarray*}

\subsubsection*{Proof of Lemma \ref{42}} 

For a description which becomes purely Eulerian at a fixed boundary across which there is an imposed velocity,
\begin{eqnarray*}
- \frac{1}{2} \rho \left< {\tilde{\bv}}, {\tilde {\bv}} \left(
{\tilde {\bF}}^{-t} {\tilde {\bN}} \cdot \left( {\tilde {\bv}} -
{\tilde {\bv}}^{\mbox{\scriptsize {\em ref}}} \right) \right) {\tilde
J} \right>_{L^2(\tilde \Gamma)} &=& - \frac{1}{2} \rho \left< {\bv},
{\bv} \left({\bn} \cdot {\bv} \right) \right>_{L^2(\Gamma)} \\
&\le& \frac{1}{2} \rho \left| \int_{\Gamma} \left[ {\bv}\cdot{\bv}
({\bn}\cdot{\bv})^{\frac{1}{2}}\right]
({\bn}\cdot{\bv})^{\frac{1}{2}} \ d \Gamma \right| \\
&\le& \frac{1}{2} \rho \left( \int_{\Gamma} {\bn} \cdot {\bv} \ d \Gamma \right)^{\frac{1}{2}} \left( \int_{\Gamma}
\left({\bv}\cdot{\bv}\right)^2 {\bn}\cdot{\bv} \ d \Gamma
\right)^{\frac{1}{2}} \\
&& \hspace{15mm} \mbox{\it (by Schwarz inequality)} \\
&=& 0
\end{eqnarray*}
since the total flow across the boundaries, $\displaystyle \int_{\Gamma} {\bn} \cdot {\bv} \ d \Gamma$ of a fixed volume of incompressible fluid must be zero. 

\subsubsection*{Proof of Lemma \ref{354}} 

By equation (\ref{309}) of the Lemma \ref{355} proof. 

\subsubsection*{Proof of Lemma \ref{57}} 

By Young's inequality
\begin{eqnarray} \label{51}
\left|\left| {\tilde {\bv}}_{n+1} {\tilde J}_{n +
\alpha}^{\frac{1}{2}}  \right|\right|_{L^2({\tilde \Omega}_{n +
\alpha})} \ \left|\left| {\tilde {\bv}}_{n} {\tilde J}_{n +
\alpha}^{\frac{1}{2}}  \right|\right|_{L^2({\tilde \Omega}_{n +
\alpha})} &\le& \left(\frac{c}{2}\right) \left|\left| {\tilde
{\bv}}_{n+1} {\tilde J}_{n + \alpha}^{\frac{1}{2}}
\right|\right|_{L^2({\tilde \Omega}_{n + \alpha})}^2 \nonumber \\
&& + \left( \frac{1}{2c} \right) \left|\left| {\tilde {\bv}}_{n}
{\tilde J}_{n + \alpha}^{\frac{1}{2}}  \right|\right|_{L^2({\tilde
\Omega}_{n + \alpha})}^2
\end{eqnarray}
for $c > 0$. Writing ${\tilde K}({\tilde {\bv}}_{n+\alpha})$
explicitly in terms of the ``intermediate'' velocity definition,
(\ref{52}), leads to
\begin{eqnarray*}
{\tilde K}({\tilde {\bv}}_{n + \alpha}) &=& \alpha^2 {\tilde K}({\tilde
{\bv}}_{n  + 1}) + (1 - \alpha)^2 {\tilde K}({\tilde {\bv}}_{n }) + 2
\alpha(1-\alpha) \left< {\tilde {\bv}}_{n  + 1},{\tilde {\bv}}_{n}
{\tilde J}_{n + \alpha} \right>_{L^2({\tilde \Omega}_{n + \alpha})} \\
&\ge& \alpha^2 {\tilde K}({\tilde {\bv}}_{n + 1}) + (1 - \alpha)^2
{\tilde K}({\tilde {\bv}}_{n }) \\
&& - 2 \alpha(1-\alpha)\left|\left| {\tilde {\bv}}_{n + 1} {\tilde
J}_{n + \alpha}^{\frac{1}{2}} \right|\right|_{L^2({\tilde \Omega}_{n +
\alpha})} \ \left|\left| {\tilde {\bv}}_{n} {\tilde J}_{n +
\alpha}^{\frac{1}{2}}  \right|\right|_{L^2({\tilde \Omega}_{n +
\alpha})} \\
&\ge& \alpha \left[ \alpha - (1 - \alpha) c \right] {\tilde K}({\tilde
{\bv}}_{n  + 1}) + (1-\alpha) \left[ (1-\alpha) - \frac{\alpha}{c}
\right] {\tilde K}({\tilde {\bv}}_{n })
\end{eqnarray*}
using equation (\ref{51}). 

\bibliography{G98-37}

\end{document}